%% file: ms_revised3.tex
\documentclass[12pt,preprint]{aastex}
\def\SP{\let\\=\empty\futurelet\C\space }
\def\etal{et\SP al.\SP }

\def\kms{kms$^{-1}$}
\def\h-1{$h^{-1}$}

\def\void#1{{}}

\def\h1{$h^{-1}$}

\def\kms{km~s$^{-1}$ }
\def\etal{et al.\,}
\def\ie{i.e.\,}
\def\eg{e.g., \,}

\def\lsim{~\rlap{$<$}{\lower 1.0ex\hbox{$\sim$}}}
\def\gsim{~\rlap{$>$}{\lower 1.0ex\hbox{$\sim$}}}

\def\mg{Mg$_b$\SP}
\def\fe{$<{\rm{Fe}}>$\SP}
\begin{document}

\title{Ages and Metallicities of Early-Type Void Galaxies from Line
Strength Measurements }


\author{
Gary Wegner\altaffilmark{\ref{Hanover}},
Norman A. Grogin\altaffilmark{\ref{JHU}}
}

\newcounter{address}
\setcounter{address}{1}
\altaffiltext{\theaddress}{\stepcounter{address}
Department of Physics \& Astronomy, Dartmouth College, 6127 Wilder
Laboratory, Hanover, NH 03755; gaw@bellz.dartmouth.edu\label{Hanover}}

\altaffiltext{\theaddress}{\stepcounter{address}
School of Earth and Space Exploration, Arizona State University,
PO Box 871404, Tempe, AZ 85287-1404; nagrogin@asu.edu
\label{JHU}}


\begin{abstract}

We present spectroscopic observations of 26 galaxies of type
E and S0, based on their blue morphologies, 
located in voids by the study of Grogin \& Geller (1999). 
Measurements of redshift, velocity dispersion, and four
Lick line indices, \mg, Fe5270, Fe5335, and H$\beta$ with their errors
are given for all of these galaxies, 
along with H$\beta$, [OIII], H$\alpha$, and [NII]
emission line strengths for a subset of these objects. 
These sources are brighter than $M^*$ for low density regions
and tend to be bluer than their counterpart
early-type objects in high density regions. 
Using the models of Thomas \etal (2003) gives 
metal abundances and ages with a median
$\alpha$ enhancement, $[\alpha/Fe]$ = +0.13, 
and median metals abundance, 
$[Z/H] = +0.22$, values comparable to
those found for E and S0 galaxies in clusters,
but with a wider spread in [Z/H] towards low values. 
If the emission line subsample is interpreted as younger, 
the proportion of young objects is higher than for early-types in
higher density regions.
There is a significant incidence of sources in the sample with
emission lines in their spectra (46\% with 
H$\beta$ and [OIII] and 69\% with H$\alpha$ or [NII]) as well as shells 
and rings in their morphologies (19\%). The diagnostic 
$\log[\rm{NII}]/\rm{H}\alpha$, 
$\log[\rm{OIII}]/\rm{H}\beta$ diagram places 10 of 12 emission line
galaxies in or near the star forming and liner region 
and two among the Seyferts.
The H$\alpha$ fluxes indicate star formation rates of 
0.2 to 1.0 $M_{\odot}\rm{yr^{-1}}$. 
The percentage of these early-type
void galaxies undergoing star formation appears to be higher 
compared to their cluster counterparts and the range of ages wider.

\end{abstract}


\section{INTRODUCTION}
Less has been known about early-type galaxies in low density environments 
than those in the higher density regions of galaxy clusters 
outlining large-scale structure.
Cosmological data (\eg Spergel \etal 2003; Efstathiou \etal 2002) 
and many recent observations of galaxies and galaxy
clusters indicate that galaxy evolution follows a complicated
tree of mergers (\eg Cole \etal 2000; Benson \etal 2001).
Galaxy  color indices and line strengths have been brought to bear on these
questions using increasingly complicated models
(\eg Tinsley 1978; Bruzual \& Charlot 1993; Worthey 1994; Colless \etal 1999; 
Nelan \etal 2005). Peebles (2001) and Nusser, Gubser, \& Peebles (2005) 
discusses problems posed by the void phenomenon, which 
offer a clue to understanding large-scale structure formation. Recent
progress in understanding galaxy evolution has given impetus
to considering galaxies in the voids.  
The question then is, how do the galaxies in the voids differ from those in 
clusters?

For void galaxies (hereafter VGs), the problem that arises is the construction
of a sizable sample of objects definitely in the voids.  Consequently
there exists a range of definitions of VGs. Grogin \&
Geller (1999, hereafter GG99) located galaxies within three
nearby voids using a 5$h^{-1}$ Mpc-smoothed density estimator and the
CfA2 redshift survey \cite{gh89} of the Updated Zwicky Catalog (UZC,
Falco \etal 1989). In general, GG99 concluded what is now recognized 
by many studies, viz. compared to galaxies in
higher density regions, the $(B-R)$ color distribution of VGs is
significantly bluer and the most luminous galaxies ($M_R \leq
-21$ mag.) are absent. Grogin \& Geller (2000; hereafter GG00) noted that 
while 
galaxy morphologies vary little with density, there is a reduction in the
frequency of early-types and and an increase of irr/pec types in the
lower density environments. G00 also found that neighbors in the voids have
1) sharply reduced velocity separations and 2) much higher incidence of 
prolific star formation as compared to near-neighbors outside voids.
Patiri \etal (2006) find similar results from
495 SDSS galaxies. 

Numerous investigations have studied galaxy properties in
different density environments. Bernardi \etal (1998) 
broke the ENEAR early-type galaxies into
those belonging to clusters and those in the field. They
used the \mg index and concluded that most of the stars 
originated  at $z > 3$ with no 
difference between high and low density regions.
Kuntschner \etal (2002) studied 9 early-type galaxies in low density 
environments and concluded that their 
formation continued to $z \leq 1$ while in clusters it ended by 
$z \geq 2$. Colbert \etal (2001) found 30 isolated early-type
galaxies using the RC3 (de Vaucouleurs \etal 1991). 
Thomas \etal (2005) compared 54 early-type galaxies with 70 in low
density regions, finding evidence for the influence of environment on
the stellar populations. Their massive objects in low density regions
appeared about 2 Gyr younger and marginally more metal rich with no
significant $[\alpha/Fe]$ differences.
Colobert \etal (2006)
have measured Lick indices for 22 objects in low density regions,
concluding that
they show a wider range in metals and ages than cluster galaxies. 
Sil'chenko (2006) studied nearby S0s in different environments and finds
objects in sparse environments tend to be younger than those in dense regions.

The number of objects used in statistical studies has been steadily increasing.
Rojas \etal (2005) used 1010 galaxies
in voids, but these are not sorted by their morphological types.
Bernardi \etal (2003a,b) studied the
properties of $\sim$9000 early-type galaxies using spectroscopic data from
the Sloan Digital Sky Survey (SDSS); although the Fundamental Plane of
early-types in dense and sparse regions may vary slightly, 
the spectra and color-magnitude relations were not seen
to differ significantly.
Bernardi \etal (2006) extended studies 
to 39320 early-type galaxies in different environments finding
that objects in low density regions have had the more recent 
star formation. Kauffmann \etal (2006) studied star formation in 46892
SDSS galaxies in different environments and determine that the most
sensitive environmental differences occur for objects with stellar mass
$M_{\star} < 3\times10^{10}M_{\odot}$. One should note that although numerous,
the SDSS spectra are fiber-fed and thus sample only the inner regions of
nearby galaxies; even as far as $\sim$150 Mpc, a 3$\arcsec$ diameter aperture
corresponds to $\sim$2 kpc on the object.

Similar data are becoming available for high redshift galaxies (\eg  
Faber \etal 2005; Glazebrook
\etal 2004); these latter authors conclude that there is evidence that galaxy
formation has continued up to the present and that at least 67\% by
mass have formed since $z \approx 2$. Schiavon \etal (2006) find ages
in red field galaxies are inconsistent with all their stars having formed
at high redshift and then passively evolving.    

Other investigators studied galaxies in 
different density regions of clusters. Poggianti et al. (2001a,b) analyzed 
Lick indices for Coma galaxies finding trends with both galaxy luminosity 
and environment. Carter et al. (2002) saw a significant $Mg_2$ radial gradient
in Coma, but only weak $<Fe>$ and H$\beta$ gradients.
Edwards \etal (2002) studied  
giant and dwarf galaxies in Coma, obtained evidence for differences 
in dynamics, and concluded that the Coma cluster was 
formed by mergers. Mehlert et al. (2003) also surmised that mergers 
of galaxies occurred from the
radial dependence of line strengths in individual Coma galaxies.
Ogando \etal (2005) utilized the Mg$_2$ radial gradients in a much larger
field sample and inferred that both the monolithic dissipative collapse and 
hierarchical merging processes can contribute to the buildup of
elliptical galaxies. 
Nelan \etal (2005) studied 4097 red sequence galaxies in 93 low $z$ clusters
using Lick indices and velocity dispersions and found evidence for
evolutionary effects including 'downsizing'~(\ie~stars in
the most massive ellipticals have existed on the red sequence since
early epochs, but most lower mass ellipticals have only recently reached
the red sequence). Smith \etal (2004, 2006) utilized
this same galaxy sample and found radial age and [$\alpha/$Fe] 
gradients within galaxy clusters, suggesting 
that early-type galaxies in the cluster cores were
accreted earlier than those at the outer radii. Rines \etal (2005) have used
H$\alpha$ emission to estimate star formation in cluster galaxies as a
function of environment and found that the kinematics of star forming 
galaxies in the infall regions of clusters rule out some 'splash back'
mechanisms.

Whilst earlier semi-analytical models of hierarchical 
galaxy formation had problems matching the
observed properties of elliptical galaxies
including `downsizing,' more recent models 
are more successful. Tantalo \&
Chiosi (2002) used N-body Tree-SPH models of elliptical galaxies
in normal environments. They found that the
hierarchical scenario enhances [$\alpha$/Fe] over a wide range of ages.
De Lucia \etal (2005) studied the evolutionary histories of ellipticals
in the hierarchical model 
using large high-resolution simulations combined withe a semi-analytic
model in the concordance 
$\Lambda$CDM cosmology. They predict that the average ages, metallicites,
and redness of elliptical galaxies 
increase with the density of their environment and 
achieve better agreement with the observations than older models.
Utilizing models that include the AGN feedback mechanism of Croton \etal
(2006) which suppresses gas condensation by cooling flows, they found that star
formation peaks near $z = 5$, but lasts longer for
field ellipticals than for the cluster ellipticals where it is more
sharply peaked for galaxies of mass 10$^{12}$M$_\odot$ than for ones
of 10$^9$M$_\odot$.  
Hopkins \etal (2006) compute merger models which include super massive
black holes and match well 
several of the existing data for early-type galaxies.
Khochfar \& Silk (2006) argue that stars in ellipticals and bulges form
from two main components, initially from mergers which are then added to
later by quiescent gaseous disks formed from mergers.

In this paper we report spectroscopic observations of the 
Lick system Mg, Fe, and H$\beta$ line
strengths as well as emission line strengths
for 26 galaxies located in nearby voids selected using the
rigorous method of GG99 and GG00.   
\S 2 presents the sample selection and its basic properties.
\S 3 describes the observations and gives the line strength
measurements. \S 4 uses the line indices to determine metal 
abundances and ages. \S 5
compares these data with galaxies in denser
regions and models and determines star formation rates. \S 6
presents our conclusions and discusses them in light of current
ideas on galaxy formation. Throughout this paper we
assume $H_0 = 70$ \kms, $\Omega_M = 0.3$, and $\Omega_{\Lambda} =0.7$.

\section{SAMPLE DEFINITION}
\subsection{Void Locations}
Figures 1 and 2
give the locations of the 26 galaxies in the present study.
They are a subset of the VGs 
defined in the GG99 sample, where the large-scale galaxy number density was
estimated by smoothing the CfA2 galaxy distribution with a
$5h^{-1}$~Mpc Gaussian kernel normalized by the CfA2 selection
function.  From this large-scale density map, GG99 targeted 280
galaxies within and bordering three prominent ``voids'' in CfA2 at
$cz\sim5000$--10000~\kms (see Fig.~1 of GG99).  Fainter galaxies in
the same regions as found by the narrower but deeper Century Survey
($R\lesssim16.1$, Geller et al.~1997; Wegner \etal 2001) and 15R 
Survey ($R\lesssim15.4$,
Geller et al., in prep.) were also included (see Figs.~1 and 2 of
GG99) in order to investigate lower-luminosity VGs below the UZC flux
limit ($B\lesssim 15.5$).

We selected early-type VGs from the GG99 sample according to the
$B$-band morphological classifications (GG00) of the GG99 imaging,
taken with $\approx1\farcs5$--2\farcs0 seeing.  We secured
spectroscopic observations for 26 of the 35 galaxies in
GG99 located in under dense regions (density contrast $n/\bar n<1$) 
and with revised
Hubble type $T\leq-2$ (S0 or earlier), where $n$ is the smoothed galaxy 
number density at a point ($\alpha, \delta, z$) and $\overline{n}$ 
is the mean density for the CfA2 survey as defined in GG99. 

The 26 spectroscopically observed galaxies are
listed in Table 1.  Column (1)
is an ordinal reference number, and column (2) is the redshift-catalog
designation.  Columns (3) and (4) list the J2000 right ascension and
declination, respectively. 
Column (5) is the redshift from GG99 and GG00 corrected to the local standard
of rest using $300~ \rm{sin}~ l~ \rm{cos}~ b$ \kms. 
Column (6) gives new
morphological types estimated using the blue copies of the POSS2
described in \S 2.2. Finally,
Columns (7), (8), and (9) give the $B$, $R$, and $K$-band absolute magnitudes.
The $B$ and $R$ data are described in GG99 and GG00 and 
all the VGs were observed by 2Mass (Skrutskie \etal 2006). 
We used the total apparent magnitudes and the redshifts to calculate the 
absolute magnitudes corrected for galactic absorption using
Schlegel et al. (1998) in NED. 

\subsection{Morphologies and Luminosities}
Histograms of the absolute magnitudes of the 26 observed VGs
are given in Figure 3. 
Comparing to luminosity functions of nearby galaxy clusters \eg
Virgo and Fornax (\eg Jerjen \& Tammann 1997) or more recent redshift
survey results shows that these VGs are at the bright end
of the luminosity function for ellipticals. 
They have an average absolute magnitude $\overline{M}_B =-18.7$,
$\overline{M}_R =-20.4$, and $\overline{M}_K =-23.5$.
Thus most of our VGs are brighter than $M^*_{bj} \approx -17.8$
found for early-type void galaxies by Croton \etal (2005) or 
$M^*_r \approx -19.0$ (Hoyle \etal 2005) which is roughly a magnitude fainter
than for objects in higher density regions.. 

The color-magnitude diagram for  the VGs is also shown in Figure 3. For 
comparison, the boundaries of the Coma cluster red sequence 
estimated from Mobasher et al. (2001) 
assuming a distance modulus, $DM = 35.0$ mag. are given.
The present VGs mostly lie near the Coma red sequence, but four
objects definitely lie below (i.e. are bluer) of which three have 
the H$\beta$ emission 
discussed in \S 3.2.2. The  average $(B-R)$ of those with
emission is $(B - R) = 1.3$ compared to 1.4 without emission respectively. 

Figure 4 presents $2^\prime \times 2^\prime$ pictures of the 26 
observed VGs
taken from the digitized blue POSS-II \footnote{ 
The Digitized Sky Surveys were produced at the Space Telescope Science 
Institute under U.S. Government grant NAG W-2166. The images of these surveys 
are based on photographic data obtained using the Oschin Schmidt Telescope on 
Palomar Mountain and the UK Schmidt Telescope. The plates were processed into 
the present compressed digital form with the permission of these institutions.
The Second Palomar Observatory Sky Survey (POSS-II) was made by the California 
Institute of Technology with funds from the National Science Foundation, the 
National Geographic Society, the Sloan Foundation, the Samuel Oschin 
Foundation, and the Eastman Kodak Corporation.
The Oschin Schmidt Telescope is operated by the California Institute of 
Technology and Palomar Observatory. 
}.
Blue images have the
advantage that they are more sensitive to spiral structure and star forming
regions. Although most of these objects
appear as early-type (E or S0) galaxies, five or 19\%
show evidence of structure \eg rings or
shells and deep imaging may increase this proportion. 
Colbert \etal (2001) also examined morphologies of 30 
isolated early-type galaxies using deep imaging
and found a 41\% incidence of shells or tidal features compared
to 8\% for a group sample. 

\section {OBSERVATIONS}
\subsection{Spectroscopic Data}
Long slit spectra of
the 26 VGs were observed 1999 November and  2000 May, at 
the MDM Observatory on Kitt Peak using the MODSPEC spectrograph attached to
the 2.4m Hiltner telescope.
Additional observations of 14 VGs were
secured in 2003 November and 2004 April with the same instrumental setup.
The detector was a $2048^2$ Site device 
covering the wavelength region of $\lambda\lambda$ 4900 to 5800. 
A 1\farcs9 slit was employed with a 1200 lines 
mm$^{-1}$ grating (original
dispersion of 1.01 \AA pix$^{-1}$)
giving 2.0~\AA~resolution. 
For all spectra, the slit length exceeded 6$\arcmin$ which 
can be compared with 
the $\sim$1$\arcmin$ size of the galaxies. Generally each
observation consisted of three 30 or 60 minute integrations
which yielded a combined $S/N > 60 -90$ per pixel near $\lambda$ 5200.
The slit ran N-S for the 1999-2000 observations; in 2003-2004 it was
along the galaxy's major axis. Three kinds of 
standards were observed: spectrophotometric
as implemented in IRAF\footnote{IRAF is distributed 
by the National Optical Astronomy Observatories which is operated by
the Association of Universities for Research in Astronomy, Inc. under
contract with the National Science Foundation.}; 
radial velocity with
spectral class near K III; and Lick Standards from 
Worthey \etal (1994) and Trager \etal (1998) for
line strength calibration. Hg-Ne-Xe comparisons
were taken before and after object exposures. Typically about five Lick and
radial velocity standards and four flux standards were observed each night. 

Additional spectra of 22 of the VGs covering $\lambda\lambda 6260 - 7175$
and $\lambda\lambda 4250 - 5150$
were secured in 2006 November using the CCDS spectrograph with a
Loral $1200 \times 800$ CCD on the Hiltner
telescope. A 600 lines mm$^{-1}$ grating (original dispersion 0.79 
\AA pix$^{-1}$), a LB370 
order separation filter, and a 1\farcs7 slit placed N-S
were used which yields a 3.0~\AA~resolution. Two 15 min.
exposures were obtained of each galaxy giving $S/N \sim 20-40$ in the 
continuum 
near $\lambda$6700 when summed, but the emission lines were 
typically a factor 30 -100
stronger. Spectra of eight objects, including the six remaining,
were obtained in 2007 March with the 
same spectrograph but with a 150 lines mm$^{-1}$ grating
(original dispersion 3.29 \AA pix$^{-1}$)
and a 1\farcs3 slit running E-W yielding 12 \AA~resolution covering the
wavelength range $\lambda\lambda 4000 -7200$. Single 15 min. exposures of the
galaxies were taken with this setup which yielded $S/N \sim 30$ in the
continuum near $\lambda$6700 but again the emission lines can be many times
stronger than the continuum.
Calibration spectra and stars were observed as above. 

Standard reductions of the 
spectra used the {\it longslit} option in IRAF. This
included bias subtraction, flat fielding,
and wavelength calibration. For sky
subtraction the {\it background} routine was employed. The
final one dimensional spectra were extracted using {\it apsum} with the 
variance weighting option and including all of the galaxy's light 
to a point where the luminosity had fallen to 10\% of its peak value.

Figure 5 gives examples of the blue and red portions of the 
VG spectra. Note that although all have been corrected for the
spectral response of the apparatus, the intensity scales are on
a relative scale. Galaxies 21 and 22
show the strong emission lines in both the blue and red regions. If these
lines are eliminated, the underlying spectra show the typical \mg and
iron lines although a bluer continuum. Objects 15 and 17 have weaker 
emission lines
in the blue and have absorption lines 
that are more typical of early-types, but they have strong 
emission in the red. Galaxies 4 and 24 have emission in the red, primarily
weaker [NII]$\lambda6583$, weak or absent H$\alpha$ presumably
filling in the underlying absorption line, and 
no detectable blue emissions. Many of
other galaxies, \eg 3, 5, and 18 (not shown) appear as normal early-types
with no detectable emission lines. The emission lines are discussed
further in \S 5.2.

The heliocentric redshifts, $cz$, and velocity dispersions, $\sigma$,  
were measured  from the higher resolution 1999 - 2004 spectra as
in Wegner \etal (1999, 2003) employing  the cross correlation method
(Tonry \& Davis 1979) in the IRAF task {\it fxcor}. The  
correlation peak width is calibrated by convolving the 
standard star spectra with
Gaussians of known width to obtain $\sigma$; measurements
used rest wavelengths of $\lambda\lambda$ 4900 to 5500 in the
galaxy spectra. 

\subsection{Line Strength Measurements}
\subsubsection{Absorption Line Data}
Line strengths for four Lick system absorption features were obtained for all
spectra taken in 1999 - 2004: H$\beta$, 
\mg, Fe$\lambda$5270, and Fe$\lambda$5335. 
The galaxy and standard star spectra were Gaussian convolved to degrade
their resolution to the Lick 8.4 \AA\ value and corrected for instrumental 
spectral sensitivity using spectrophotometric standards. 
Instrumental indices were placed on the Lick system utilizing
the stellar standards and applied to the galaxy data. 
The line indices are defined in Trager \etal 
(1998). Figure 6 compares our standard star indices 
with those published (Worthey \etal (1994); 
Trager \etal 1998) after correction: the H$\beta$, \mg, Fe$\lambda$5270, and
Fe$\lambda$5335 were multiplied by 0.997, 1.393, 1.284, and 1.862 
respectively.
Galaxy indices were corrected for 
the broadening due to $\sigma$. Several authors
including Gonzales (1993) and Trager 
\etal (1998) have published nearly identical corrections for many
indices. Here we adopt Gonzales' (1993) values, although 
we note that these can
change for stellar populations not well matched by K giant spectra
(Schiavon 2007).
The errors in the redshifts, velocity dispersions, and 
line indices were estimated from the standard
deviation of the mean of individual spectra of each galaxy. 

\subsubsection{Emission Line Data and H$\beta$ Corrections}

Equivalent widths for the emission lines were determined
using the IRAF routine $splot$. 
These data are discussed further in \S 5.2.
Most of these objects have H$\alpha$ emission measurements that are
reported in GG00. On the whole the agreement is good, but with some
large differences which we attribute as due to the range in slit widths
and position angles used in the different observations and could also
affect the simple H$\alpha$ infill correction used below. 

It is well known that H$\beta$ indices can suffer from 
emission contamination  Many investigators 
corrected for this emission by subtracting a fraction of the combined 
equivalent widths of the $\lambda$5007 and $\lambda$4957 [OIII] lines.
Gonzales (1993) multiplies by 0.7 while Kuntschner \etal (2001)
determine 0.6. Nelan \etal (2005) show that this simple correction
is subject to large errors and often underestimates the H$\beta$
emission by large factors. However they demonstrate a tight correlation between
H$\alpha$ and H$\beta$ finding that the H$\alpha$/H$\beta$ ratio
is 4.5$^{+1.5}_{-1.6}$. This can be compared with Stasi\'nska
\etal (2004) who demonstrated a tight relation for spirals and obtained
a ratio of 5.4. Here we utilize the 4.5 factor of Nelan \etal to correct
H$\beta$.

H$\alpha$ must also be adjusted for underlying absorption
in the lower resolution spectra shown in Figure 4.
The mean of the four strongest H$\alpha$ absorption line
equivalent widths in the VGs which appear to be free of any emission lines
is $2.32 \pm 0.22$ \AA. We then compensate for absorption 
assuming this mean value.
We do not make an absorption correction for H$\beta$ which is observed in
the higher resolution spectra and where the line profiles are well
resolved. Here negative values of line strengths denote emission features.
Therefore the corrections to the equivalent widths of the two lines in 
\AA~were obtained using the following two steps:
\begin{enumerate}
\item Correct the equivalent width of 
H$\alpha$ for absorption using $\rm{H}\alpha^{(em)} = 
\rm{H}\alpha^{(obs)} -2.32$.
\item The emission correction for H$\beta$ is then obtained using 
$\rm{Corr}(\rm{H}\beta) = \rm{H}\alpha^{(em)}/4.5$.
\end{enumerate}

\subsubsection{Line Strength Measurement Results}

Table 2 summarizes the absorption line data for the VGs.
Column (1) lists the ID number from Table 1. 
Column (2) is the number of runs the galaxy was observed.
Columns (3) and (4) give our redshift and
its error. Columns (5) and (6) are $\sigma$ and its error. Columns (7) and (8)
are the H$\beta$ index, uncorrected for emission,
 and its error; column (9) is the correction to H$\beta$
as derived above Columns (10)--(15) are 
the \mg and iron indices and their errors. No aperture corrections have been
applied to the values in the table. Using the correction methods of
J{\o}rgensen \etal (1995) and Nelan \etal (2005), these are found to be
of order 0.01.

Table 3 presents the equivalent width measurements of H$\alpha$, 
[NII]$\lambda$6583, H$\beta$, and [OIII]$\lambda5007$. Column (1) is the
ID number, Columns (2) and (3) give the uncorrected equivalent widths of the
first two lines. Column (3) is the value of $\log[\rm{NII}]/\rm{H}\alpha$ 
using H$\alpha$ corrected as above. Columns (4) and (5) are the uncorrected
equivalent widths of the second two lines and column (6) is
$\log[\rm{OIII}]/\rm{H}\beta$ using uncorrected H$\beta$. Emission
lines have negative values and absorption lines are positive.

\section{DERIVED METAL ABUNDANCES AND AGES }
Below we use the combined indices defined as follows:

$${{<\rm{Fe}>} = {(Fe5270+Fe5335)/2}}$$

$${\rm [MgFe]'} = \sqrt{({\rm Mg_b\times}[0.72\times {\rm Fe}5270 + 0.28\times 
{\rm Fe}5335])}  $$ 

With our 26 galaxy sample and the four measured line indices, we did not
carry out a simultaneous multidimensional fit to  
determine the ages and metal abundances
(e.g. Nelan et al. 2005). 
As a first approximation, we estimated [$\alpha$/Fe] and
[Z/H] for the VGs using the grid of models by Thomas \etal (2003; 
hereafter TMB) assuming an age of $12\times10^9$ years for
all objects. This gives a 
reasonable mean value for [$\alpha$/Fe] but not [Z/H] 
which depends more
strongly on age. As a second step, 
we subdivided the sample by age and redetermined the
metal abundances using the mean [$\alpha$/Fe].
These results are discussed in \S5. 

\subsection{Alpha and Metal Element Abundances}
Figure 7 shows the observed \mg and \fe indices with error bars 
from Table 2 plotted on the TMB grid of 
constant [Z/H] and [$\alpha$/Fe] for $12\times10^9$ year models. 
Whilst the galaxies widely
span [Z/H], most lie within [$\alpha/\rm{Fe}] = 0.0$ 
to 0.3 and lie near a mean of about 0.15. 
Compared with similar plots for cluster galaxies (\eg Coma; Mehlert \etal
2003), Thomas \etal (2005), and Bernardi \etal (2006), higher [Z/H] values 
corresponding to the upper right
portion of the figure are typical, but a tail of 
objects fall into the lower left region where
eight, mostly emission line galaxies, have apparent
[Z/H] from about -0.33 to below -2.25.
Normal ellipticals do not occupy this region of the
diagram, but such low values [Z/H]
are found for many globular clusters which 
also lie in this portion of Figure 7 (Trager \etal 1998). 
However, although the globular clusters are known to be 
old, an alternative interpretation of the eight VGs is
that they are young and have higher [Z/H].

Initial estimates of [$\alpha$/Fe] were 
obtained following Figure 4 in TMB using 
$(\frac{Mg_b}{<Fe>})$ which they 
demonstrate is insensitive to age and metal content. 
This method yields a median of 0.15 and a mean of 
$\overline{[\alpha/Fe]} = 0.16\pm 0.03$ for our VGs excluding the two most 
negative values. This result is consistent with cluster early-type galaxies
(\eg Nelan \etal 2005) and as shown in \S 4.3, is little
affected by the choice of age.

Such age insensitivity does not hold for [Z/H] because of how curves of
constant [Z/H] shift in the \fe - \mg diagram. Qualitatively, a
given galaxy placed in a 
\fe - \mg diagram of fixed age, such as Figure 7,
has higher [Z/H] using younger models. Also the young metal
poor objects are
most strongly affected. This can be seen from the following two
examples: Object 22 has [Z/H] = -2.3 assuming an age
of $12\times10^9$ years, but [Z/H] = -1.1 if it is $1\times10^9$
years old. Object 2, apparently older and having higher metals, has
[Z/H] = 0.00 if it is $12\times10^9$ years old,
but using $15\times10^9$ years only changes [Z/H] to -0.03. 
Therefore objects that are of low age and [Z/H] the most sensitive
to changes in model age.

\subsection{Age Estimates}

\subsubsection{The Weak or No-emission Line Galaxies}
Ages, $\tau$, for most of the VGs could be obtained next using
Figure 8 and data in Table 2.
Fortunately age depends
weakly on [Z/H] if [$\alpha$/Fe] is known and
the grid was interpolated from TMB employing $[\alpha/Fe] = 0.15$.
Twenty of the galaxies are covered by the TMB models and yield
consistent ages and metallicities although nine of these have
some detectable emission in Table 3. 

We stress that in practice 
using H$\beta$ makes this problematical due to the emission, 
for in Figure 8 and
the ${\rm [MgFe]'}$ - H$\beta$ plot, H$\beta$ is essentially the  
age indicator.
Our data do not currently cover H$\gamma$ which suffers less 
from emission ( See \eg Schiavon 2007; 
Bernardi \etal 2006; Poggianti \etal 2001a for
discussions.) and tends to deliver ages lower by $\sim10^9$ years.
Nevertheless, the H$\beta$ corrections are small and the ages are
relatively unaffected.

\subsubsection{The Strong Emission Line Galaxies}
Six of the galaxies (1, 12, 21, 22, 25, and 26) lie outside the grid in 
Figure 8 and require comment as their ages cannot be constrained using 
this diagram.

Objects (1 and 12) lie just below
the 15 Gyr isochrone in Figure 8. Their spectra show the strong 
features typical of early-type galaxies including \mg, NaD, and
Fe lines. They both require H$\beta$ emission corrections and 
it is plausible that these corrections are too
small and they belong with the
other objects of similar \fe in Figure 8. We have assigned these galaxies
values of $\log\tau = 10.2 \pm 0.5$. 

Objects (21 and 26), after correction for emission, still lie
far below the 15 Gyr isochrone, which naively would indicate
ages $\gg15$ Gyr. Number 21 has the strongest H$\beta$ 
emission of all our VGs 
and number 26 has the bluest $(B-R)$ color.
Although one cannot definitely determine whether these
objects are either metal poor, young, or both with the current data, 
their non-typical blue spectra suggest
that they contain young stars. In addition to the TMB models,  
Schiavan's (2007) and Bruzual \& Charlot's (2003) Lick indices
with varying metal content and age but with only  
solar $\alpha$ abundance demonstrate that age-line strength
relations depend strongly on metal content for young objects. For 
${\rm [MgFe]'}$ and Fe5270, when ${\rm [MgFe]'}$ $<$ 2, 
inferred age drops significantly for a wide abundance range and would
indicate ages $\tau < 3$ Gyr for these two galaxies.

Objects (22 and 25) also have strong emission lines and lie 
outside Figure 8, far above the other galaxies. These would also
appear to have ages, well below  3 Gyr, but cannot be constrained better.

\subsection{Final Metal Abundance Estimates}

Given the ages, $\tau$ determined from Figure 8, we have 
subdivided the 22 galaxies by age and determined 
$[\alpha/Fe]$ and [Z/H] along with estimates of their errors
using the \mg - \fe diagrams from the TMB models of age $\tau_{TMB}$:

\begin{enumerate}
\item $\tau \geq 9$ Gyr, $\tau_{TMB}$ = 12 Gyr
\item $ 9 > \tau \geq 5$ Gyr, $\tau_{TMB}$ = 7 Gyr
\item $ 5 > \tau \geq 3$ Gyr, $\tau_{TMB}$ = 4 Gyr
\item $ 3 > \tau \geq 2$ Gyr, $\tau_{TMB}$ = 2 Gyr
\item $ \tau < 2$ Gyr, $\tau_{TMB}$ = 1 Gyr
\end{enumerate} 

The metal abundances have been eliminated for the four galaxies (21, 22, 25,
and 26). These would yield [Z/H] = -0.65 to -1.74 assuming $\tau = 1$ Gyr.
As explained above, it is impossible to determine their ages using H$\beta$
and \fe with the present data. It appears that diagrams such as Figure 7
cannot be employed if the stellar populations are younger than a few Gyr.
Using Schiavon's (2007) Lick indices for single stellar populations for
solar-scaled and $\alpha$-enhanced isochrones that cover a more limited
range of metallicities, one sees that the \mg or \fe indices for these
four galaxies could be fit with models of age $\tau$ = 0.2 - 0.3 Gyr and
[Z/H] values reduced much less and closer to solar.

Figure 9 gives histograms of the derived metal abundances. For $[\alpha/Fe]$
the distribution is fairly narrow with an average
$\overline{[\alpha/Fe]} = 0.15\pm 0.09$ and a median of 0.13.
This confirms the
first estimate. The distribution of [Z/H] is somewhat broader and
shows a tail towards lower values. 
The run of [Z/H] is similar to our first approximation for the older
galaxies, but the quite metal poor ones now have higher values of [Z/H].
The mean is $\overline{[Z/H]} = 0.15 \pm 0.28$ and a median of 0.22. These data are 
considered further in \S 5.

\subsection{Summary of the Age and Metal Abundance Estimates}
Table 4 presents the derived age and abundance parameters for the 26 VGs.
Column (1) is the identification number from the other tables. Columns (2) and
(3) are the logarithm of the age in Gyr, $\log\tau$ and its error from Figure 
\ref{fehbeta}. The six galaxies located outside the range of the models
in Figure 8 as discussed in \S4.2.2 are indicated with corresponding symbols. 
Columns (4) and (5) are [Z/H]  and
its  error using the galaxy's 
position in the \mg - \fe diagrams listed above   
and columns (6) and (7) are the equivalent for
[$\alpha$/Fe]. The errors reflect the observational
uncertainties, are the average of upper and lower values, and do not 
include uncertainties in the theoretical relations.

\section{DISCUSSION OF THE VG MEASUREMENTS}
\subsection{Comparison with Cluster Galaxies}
Figure 10 shows the histogram of VG ages derived using Table 4. Galaxies
with detected H$\beta$ emission are shaded.
We place objects (1 and 12) in the $>9$ column of Figure 10.
The two galaxies (21 and 26) are
indicated in the $\gg 15$ column, but adopting  Bruzual \& 
Charlot's (2003) relation would move them into the  
$\tau < 3$ Gyr column of Figure 10 where they also appear as the 
wider hatched section.
Sources (22 and 25) also appear in the $\tau <3$ column. 

We compare the age distribution in Figure 10
with the H$\beta$ ages of Coma early-type 
galaxies given in Fig. 7 of Poggianti \etal (2001a) for their
most luminous, $R < 15$ brightness subsample.
Binning as Poggianti \etal (2001a) in four
age bins, $\tau$ (viz. $\tau < 3$, 3 - 5, 5 - 9, and $> 9$ Gyr), 
there are 7, 4, 8, and 7 VGs respectively. This 
is a higher proportion of young VG galaxies relative to
the Coma sources where the corresponding numbers
are 4, 4, 8, and 25 in their inner Coma 1 region where there is a
higher proportion of objects in the  $\tau > 9$ Gy bin. Coma
appears to show an increase in the younger bins for decreasing luminosity,
but not to the extent of the VGs.

Rejecting the 10
galaxies with the largest H$\beta$ emission corrections
(as do some authors, \eg Nelan et al. 2005; Smith \etal 2006), this
leaves a subset of sources in the unshaded bins 
and the numbers of objects become 0, 3, 8, and 5 in the same bins respectively,
raising the ages, but still averaging younger than for
the cluster objects.

Figure 11 compares age, metal abundances, and luminosities
with log$\sigma$. 
We reject the values of [Z/H] for objects (21, 22, 25, and 26)
for the reasons discussed above.
The regression relations for [Z/H], [$\alpha$/Fe], and age
for cluster early-types found in Nelan \etal (2005) 
and Fig. 14 of Bernardi \etal
(2006) are shown, as is 
the Faber-Jackson relation, ($M_B$, log$\sigma$), from Dressler \etal 
(1987) for the Coma cluster. Nelan \etal (2005) further discuss
the range of values in these relations in the literature.
The Figure 11 plots show two features:
Firstly, most of the higher $\sigma$ VGs lie near the  
mean lines for the cluster galaxies.
Secondly,
the  $\log\sigma-M_B$ relation is similar to that of Coma
and that of Kuntschner \etal (2002) for their void sample. 

Finally, using the H$\beta$ ages of the VGs, Figure \ref{metalages} shows
that most lie near the mean [$\alpha$/Fe]
relations found by Mehlert \etal (2003) for Coma cluster early types and 
Fig. 13 of Bernardi \etal (2006) 
for ellipticals in different environments. However,
in the [Z/H] diagram some  
younger objects fall below the mean relations. 

\subsection{Properties of the Emission line Objects}
\subsubsection{The diagnostic diagram for the VGs}
The presence of H$\alpha$, H$\beta$, [NII], and [OIII] emission 
in our sample indicates ongoing star formation. 
For our VGs, 12 of 26 or 46\% are objects show detectable 
H$\beta$ and [OIII]
emission lines and 18 show at least one of H$\alpha$ or [NII]$\lambda6583$.
 Nelan \etal (2005) found
$\sim$12\% of their early-type cluster galaxies had emission.
Pimbblet \etal (2006) find 20\% for their composite sample of X-ray 
bright galaxies in clusters at $z \sim 0.1$. 
Studies of the ``E + A'' galaxies (Zabludoff \etal; 1996 and 
Quintero \etal; 2004), although  not showing emission lines are 
thought to be products of recent star formation, indicate that 
in general such early-type galaxies are rarer,
currently $\sim$1\% of the general field.
The 8 VGs with no detected emission lines are more luminous than
those with detected emission. The mean values of the two types are
$\overline{M_K} = -23.9$ and $\overline{M_K} = -23.3$ respectively. 

Figure 13 plots the data in Table 3  of
$\log[\rm{NII}]/\rm{H}\alpha$ and
$\log[\rm{OIII}]/\rm{H}\beta$ 
in the diagnostic diagram which separates
H II star-forming regions from active galactic nuclei (Osterbrock 1989).
The loci dividing the galaxy classes are from
Kewley et al. (2001), Kauffmann et al. (2003), and 
Yan et al. (2006). 
The underlying Balmer absorption is required for some
of these galaxies as can be seen from the spectra of galaxies 4 and 24
in Figure 4 where [NII]$\lambda6583$ is stronger
than H$\alpha$;  this appears to result from emission
filling underlying absorption in the H$\alpha$ line. The 2.32 \AA
average equivalent width of H$\alpha$ in absorption 
was added to
H$\alpha$ as described above and moves some of the points 
left in Figure 13. 
It appears that of the 12 galaxies in Figure 13,
six objects are star-forming, two are Seyferts, three are liners,
and one borders the liners and star-forming regions.

For comparison, several investigators including
Goudfrooi (1999) and Sarzi \etal (2006), discuss the H$\alpha$ emission 
in giant
ellipticals and conclude that the liner type are the most numerous.
Smith, Lucey \& Hudson (2007) studied the occurrence of emission
among low-luminosity galaxies in the Shapley supercluster and find
that $\sim20$\% of their early-type red-sequence members show emission 
features. Although this is a lower percentage than for the VGs, their
diagnostic diagram seems to show a similar mix of types. They report that their
most luminous galaxies brighter than about $M_R = -21.5$ mag. are mostly
AGN/LINERs, while galaxies fainter than this are mostly star forming. Our
small sample appears to agree with this. 
The two Seyferts in Figure 13 
are the sample's most luminous with $\overline{M_R}$ = -21.5 mag., while the 
Seyferts plus liners have $\overline{M_R}$ = -20.2 mag. and the star 
forming galaxies give an average of $\overline{M_R}$ = -19.2 mag.

\subsubsection{Star formation rates}
To estimate star formation rates, we take
$$SFR(M_\odot yr^{-1}) = \frac{L(H\alpha)}{1.26\times10^{41} 
\rm{ergs sec^{-1}}}$$
(Kennicut 1998). 
Using flux calibrated H$\alpha$ and H$\beta$
data described above, it is possible to estimate SFR for six galaxies that
have strongest H$\alpha$ and H$\beta$ emission (numbers 15, 17, 21, 22, 23,
and 25) and for which we have photometric calibrations.
We assumed the 
intrinsic unreddened H$\alpha$/H$\beta$ 
ratio is 2.85 and the average extinction
curve to derive the extinction parameter, $c$ (Osterbrock 1989), which has
values $c = 0.2 - 0.8$. In \S 3.2.2 we took the mean observed 
H$\alpha$/H$\beta$ ratio for the whole sample to be 4.5 for the
emission correction to H$\beta$; this gives $c = 0.5$ which is close to the
mean of these six objects.
The observed H$\alpha$ emission fluxes range from
$2\times10^{-14}$ to $4\times10^{-14} \rm{ ergs sec^{-1} cm^{-2}}$. 
and thus indicate $SFR \sim 0.2~ \rm{to}~ 1.0~M_{\odot}\rm{yr^{-1}}$,
moderate values (Kewley Geller \& Jansen 2004; Salzer \etal 2005). 

Star formation rates of this order of magnitude are also observed at a
lower frequency in many early-type galaxies.
Yi \etal (2005) used the all sky GALEX near ultraviolet color-magnitude
diagram to study star formation in nearby bright early-type galaxies and
conclude that $\sim$15\% of these objects show low-level star formation.

\section{CONCLUSIONS}
Spectroscopic data are given for 26 early-type void galaxies that were 
listed in Grogin \& Geller (1999; 2000). The morphologies of our VGs 
are all E/S0. Most of them lie in the red sequence for elliptical
galaxies, belong to the brighter end of the luminosity 
function appropriate for void galaxies, and are
not intrinsically faint void objects for which
environmental differences are expected to be more pronounced. 

Based on the spectroscopy, our main observational results are: 

(1) The values of
[$\alpha$/H] lie near 0.15 which is close to 
that of cluster ellipticals. 

(2) The metal
abundances, [Z/H], found from \mg and \fe, average
close to 0.15, normal for early-types in clusters, but
some VGs seem to have lower metal content
than found in clusters but this interpretation is clouded by the
higher sensitivity of [Z/H] for the low ages..

(3) Compared to the early-type cluster galaxies studied, \eg in 
Nelan \etal (2005), the runs of
[$\alpha$/H] and [Z/H] with $\sigma$ and age for the VGs seem to be
consistent in many cases, but there are some objects that appear
to have lower ages and metals, [Z/H].

(4) Ages for the stellar populations in the VGs were
determined from the (\fe, H$\beta$) diagram and required
corrections to H$\beta$ for emission that were derived from H$\alpha$. 
The resulting ages of several of the VGs are younger than
those found by other investigators for
bright early-types in denser environments such as Coma. 
This is particularly the case for VGs 
with emission lines that comprise the majority of galaxies younger then
5 Gyr. There is a reduced percentage of old objects ($\tau > 9$ Gyr in
the distribution which indicates a range of old and young galaxies. 

(5) The occurrence of emission in the VG
spectra is high and appears to agree with the age conclusions.
In the sample, 46\% show detectable H$\beta$ lines and 
69\% have H$\alpha$, different 
from early-types in higher density environments.

(6) The positions of 12 VGs in the diagnostic diagram indicate 
a mixture of emission types, but the signature of star formation
is common with moderate values of star formation in the range 
$SFR \sim 0.2~ \rm{to}~ 1.0~M_{\odot}/\rm{yr}$.

(7) Visual inspection of blue POSS-II images find that
five or 19\% present evidence of shell or ring structures.

The early-type VGs of this study appear to differ in 
age from their cluster counterparts.
Some of these 
trends have been indicated by other investigations referenced in 
the introduction, but here we found evidence for a prevalence of
strong emission lines and younger ages.
The most massive early-type VGs, judged from their luminosities and 
velocity dispersions, tend to 
have the highest ages and $\alpha$-enrichments as do 
their cluster counterparts.

The results for metal abundances and age determinations depend on the 
applicability of the TMB models that have been employed to obtain them. 
In addition the
ages of 10 VGs depend critically on the interpretation of 
the H$\beta$ line strengths.
Although we have conducted a straightforward interpretation of the 
data using these models,
the interplay of age, $\alpha$, and Z make determination of these
quantities tricky at best for the youngest objects which
also are the most interesting; they are the most unreliable.
New generations of models which cover a wider range of variation in
element abundances and which handle multiple starbursts 
should aid in this (\eg Schiavon 2007).

Nevertheless,
there are morphological differences between the VGs and the cluster
early-types that are model independent.
These include the incidence 
of more emission lines, shells, and rings, suggesting that many 
VGs are still undergoing hierarchical assembly. These results are
unlikely to go away.   
In the context of current thinking on galaxy formation, these findings 
for the early-type VGs seem consistent with the ``downsizing'' picture 
(\eg De Lucia \etal 2006; Nelan \etal 2005; Smith \etal 2006) in which 
the star formation and evolution stop earlier for the more massive
objects and are drawn out longer in time for the lower mass ones.
The younger ages would be consistent with a slower evolution in the lower
density environments.

Studies of void evolution seem to be consistent with these results,
but indicate that a simple process of galaxy formation
is unlikely. Void structures would 
form differently from those in the higher density regions. 
Many authors including, \eg Sheth \& van de Weygaert
(2004), Colberg \etal (2005), and van de Weygaert \& van Kampen 
(1993) indicate that the large voids
arose from the mergers of smaller voids which get squeezed down as 
overdensities in the large voids as they collapse. Radial galaxy movements are
suppressed and their peculiar motions are mostly confined to run along the
void walls where most of the galaxies reside. 
This leads to VGs in large voids forming in the
wall regions of the smaller voids as they collapse and remain in the
voids. Thus the observed VGs would arise from a 'cloud-in-void' process
(Sheth \& van de Weygaert 2004) which naturally produces a lower
density environment.    

The problem of constructing an uncontaminated VG sample which
detects the fainter void interior structures remains.
More detailed mapping of the VGs kinematical and line strength structures
as well as investigation of their immediate environments would help
resolve some of these questions.

We wish to thank Drs. R. P. Saglia and J. R. Lucey who made comments on
earlier versions of this paper and the referee, 
Dr. R. P. Schiavon for careful reading 
of the manuscript and suggestions that improved the paper.
This research has made use of the NASA/IPAC Extragalactic Database (NED) 
which is operated by the Jet Propulsion Laboratory, California Institute 
of Technology, under contract with the National Aeronautics and Space 
Administration.The Digitized Sky Surveys were produced at the Space 
Telescope Science 
Institute under U.S. Government grant NAG W-2166. The images of these surveys 
are based on photographic data obtained using the Oschin Schmidt Telescope on 
Palomar Mountain and the UK Schmidt Telescope. The plates were processed into 
the present compressed digital form with the permission of these institutions.
The Second Palomar Observatory Sky Survey (POSS-II) was made by the California 
Institute of Technology with funds from the National Science Foundation, the 
National Geographic Society, the Sloan Foundation, the Samuel Oschin 
Foundation, and the Eastman Kodak Corporation.
The Oschin Schmidt Telescope is operated by the California Institute of 
Technology and Palomar Observatory.

\newpage

\begin{deluxetable}{llcccrccc}
\rotate
\tablewidth{0pt}
\label{tab1objects}
\tablenum{1}
\tabletypesize{\scriptsize}
\tablecaption{Basic Data for the Observed Void Galaxies}
\tablehead{
\colhead{ID} & 
\colhead{Name} &
\colhead{$\alpha$} &
\colhead{$\delta$} & 
\colhead{$cz$} & 
\colhead{Type} & 
\colhead{$M_B$} & 
\colhead{$M_R$}&
\colhead{$M_K$} \\
\colhead{}& 
\colhead{}&
\colhead{J2000} & 
\colhead{J2000} & 
\colhead{\kms} &  
\colhead{}&
\colhead{mag.} &
\colhead{mag,}& 
\colhead{mag.} \\
\colhead{(1)} & 
\colhead{(2)} & 
\colhead{(3)} & 
\colhead{(4)} & 
\colhead{(5)} & 
\colhead{(6)} & 
\colhead{(7)} & 
\colhead{(8)} & 
\colhead{(9)}
  }
\startdata
1 & 15R 464.015505 & 00 03 36.098 & +10 36 12.54 &  8085 & S0 &-19.00 &-20.50  & -23.84\\
2 & CGCG 0027.9+0536 & 00 30 28.637 & +05 51 39.06 &  7238 & E2 &-19.15 &-20.59  & -24.03\\
3 & NGC 476 & 01 20 19.952 & +16 01 12.40 &  6439 & S0 &-18.78 &-20.37 & -23.81\\
4 & IC 1700 & 01 25 24.652 & +14 51 52.43 &  6484 & E0(shells) &-20.12 &-21.64 & -25.01\\
5 & CGCG 0133.6+1734 & 01 36 20.281 & +17 48 58.01 &  6065 & E2 &-18.10 &-19.71  &-23.11 \\
6 & CGCG 1348.4+3517 & 13 50 38.973 & +35 02 18.09 &  6407 & E0 &-19.55 &-21.01  & -24.42\\ 
7 & CGCG 1429.5+3929 & 14 31 27.815 & +39 15 32.14 &  7038 & SB0 &-19.22 &-20.73 & -24.15\\
8 & CGCG 1433.0+2311 & 14 35 17.143 & +22 57 46.34 &  6336 & SB0(ring)&-18.54 &-19.84 & -23.06\\
9 & CGCG 1434.7+4010 & 14 36 42.686 & +39 56 37.64 &  6217 & E1 &-18.72 &-20.19 & -23.74\\
10 & CGCG 1506.2+2208 & 15 08 29.040 & +21 56 51.73 &  6259 & S0 &-18.82 & -20.42& -23.72\\
11 & 15R 328.038850 & 15 33 41.477 & +28 08 44.24 &  5772 & S0/a & -16.95&-18.35  & -21.73\\
12 & CGCG 1536.8+3707 & 15 38 38.464 & +36 57 30.18 &  5751 & S0 &-19.02&-20.37  &-23.76\\
13 & CGCG 1549.1+2551 & 15 51 13.249 & +25 42 05.11 &  6546 & S0 &-19.33 &-20.85  & -24.18\\
14 & CGCG 1552.1+3018 & 15 54 11.455 & +30 09 21.35 &  6175 & E0 &-18.30 &-19.82 & -23.39\\
15 & CGCG 1600.6+4302 & 16 02 16.663 & +42 54 59.66 &  7541 & S0(ring) &-19.77 & -20.98 & -24.21\\
16 & CGCG 1607.5+4152 & 16 09 16.555 & +41 44 43.47 &  7891 & S0 &-19.37 &-20.81 & -24.15\\
17 & CGCG 2108.7+0318 & 21 11 15.074 & +03 31 19.42 &  5342 & E3 &-18.17 & -19.51 & -22.79\\
18 & NGC 7113 & 21 42 26.650 & +12 34 07.35 &  5989 & S0 &-19.20 & -20.89& -24.59\\
19 & CGCG 2140.0+1340 & 21 42 26.896 & +13 53 58.61 &  5631 & E2 & -18.58& -20.22 &-23.63 \\
20 & 15R 518.119354 & 21 50 27.756 & +12 38 08.91 &  6850 & S0/a &-17.27 & -18.82  & -22.29\\
21 & 15R 520.029201 & 22 35 05.336 & +10 55 25.89 &  5240 & E0 &-16.76 &-18.05  & -21.50\\
22 & CGCG 2236.9+1336 & 22 39 21.911 & +13 52 54.19 &  5423 & S0(ring) &-17.93 & -18.66&-20.89 \\
23 & NGC 7509 & 23 12 21.430 & +14 36 33.41 &  5118 & E0+star &-19.36 & -21.01 & -24.37\\
24 & NGC 7550 & 23 15 16.007 & +18 57 42.49 &  5340 & E0 &-20.29 &-21.98  & -25.45\\
25 & 15R 523.044360 & 23 37 38.094 & +12 06 00.42 & 8104 &  S0(shells)&-18.00 & -18.92 &-21.74 \\
26 &15R 523.036286 & 23 44 29.634 & +11 40 48.01 & 7090 & SB0    & -18.98 & -20.38 & -23.09\\

\enddata

\end{deluxetable}

\newpage

\begin{deluxetable}{lcccrrrrrcccccc}
\tablecolumns{15}
\rotate
\tablewidth{0pt}
\label{tab2data}
\tablenum{2}
\tabletypesize{\scriptsize}
\tablecaption{The Spectroscopic Data for the Void Galaxies}
\tablehead{
\colhead{ID} &
\colhead{$N_{obs}$}&
\colhead{$cz$} &
\colhead{$\epsilon_{cz}$}&
\colhead{$\sigma$}& 
\colhead{$\epsilon_{\sigma}$} & 
\colhead{H$\beta$}& 
\colhead{$\epsilon_{H\beta}$} & 
\colhead{$Corr_{H\beta}$} & 
\colhead{$Mg_b$}  & 
\colhead{$\epsilon_{Mg_b}$} &  
\colhead{Fe5270}& 
\colhead{$\epsilon_{Fe5270}$} & 
\colhead{Fe5335}&
\colhead{$\epsilon_{Fe5335}$} \\
\colhead{}&
\colhead{} &
\colhead{\kms}&
\colhead{\kms}&
\colhead{\kms}&
\colhead{\kms}&
\colhead{ }&
\colhead{ }&
\colhead{ }&
\colhead{ }&
\colhead{ }&
\colhead{ }&
\colhead{ }&
\colhead{ }&
\colhead{ }\\
\colhead{(1)} & 
\colhead{(2)} & 
\colhead{(3)} & 
\colhead{(4)} & 
\colhead{(5)} & 
\colhead{(6)} & 
\colhead{(7)} & 
\colhead{(8)} & 
\colhead{(9)} & 
\colhead{(10)} & 
\colhead{(11)} &
\colhead{(12)} & 
\colhead{(13)} & 
\colhead{(14)}  & 
\colhead{(15)}\\
}
\startdata

\input tab2.tex

\enddata
\end{deluxetable}

\newpage

\begin{deluxetable}{lcccccc}
\tablecolumns{7}
\rotate
\tablewidth{0pt}
\label{tabedata}
\tablenum{3}
\tabletypesize{\scriptsize}
\tablecaption{The Emission Line Data for the Void Galaxies}
\tablehead{
\colhead{ID} &
\colhead{H$\alpha$}&
\colhead{[NII]$\lambda$6583}&
\colhead{log[NII]/H$\alpha$}&
\colhead{H$\beta$}&
\colhead{[OIII]$\lambda$5007} & 
\colhead{log[OIII]/H$\beta$} \\
\colhead{} &
\colhead{$\AA$} & 
\colhead{$\AA$} &
\colhead{ } & 
\colhead{$\AA$} & 
\colhead{$\AA$}  \\
\tablevspace{5pt}
\colhead{(1)} & 
\colhead{(2)} & 
\colhead{(3)} & 
\colhead{(4)} & 
\colhead{(5)} & 
\colhead{(6)} & 
\colhead{(7)} \\    
}
\startdata

\input tab3.tex

\enddata
\end{deluxetable}

\newpage

\begin{deluxetable}{llcccrc}
\tablecolumns{9}
\rotate
\tablewidth{0pt}
\label{tab1results}
\tablenum{4}
\tabletypesize{\scriptsize}
\tablecaption{Age and Abundance Results for the Observed Void Galaxies}
\tablehead{
\colhead{ID} & 
\colhead{$\log\tau$}&
\colhead{$\epsilon_{\tau}$} &
\colhead{[Z/H]}& 
\colhead{$\epsilon_{[Z/H]}$} & 
\colhead{$[\alpha/Fe]$} &
\colhead{$\epsilon_{[\alpha/Fe]}$}  \\
\colhead{(1)} & 
\colhead{(2)} & 
\colhead{(3)} & 
\colhead{(4)} & 
\colhead{(5)} & 
\colhead{(6)} & 
\colhead{(7)} \\
  }
\startdata

01&  10.2&  0.5&  0.19&  0.09&  0.14&  0.08\\
 02& 10.1&  0.2&  0.01&  0.06&  0.11&  0.07\\
 03&  9.8&  0.3&  0.04&  0.04&  0.07&  0.04\\
 04&  9.6&  0.2&  0.31&  0.08&  0.09&  0.08\\
 05&  9.7&  0.2&  0.17&  0.05&  0.06&  0.08\\
 06&  9.9&  0.3&  0.24&  0.09&  0.10&  0.09\\
 07& 10.0&  0.3&  0.44&  0.09&  0.19&  0.10\\
 08&  9.9&  0.3& -0.27&  0.07&  0.23&  0.14\\
 09&  9.3&  0.3&  0.39&  0.18&  0.12&  0.13\\
10&  9.9&  0.3&  0.44&  0.10&  0.18&  0.10\\
11&  9.6&  0.3& -0.65&  0.27&  0.24&  0.34\\
12& 10.2&  0.5& -0.12&  0.04&  0.08&  0.06\\
13&  9.6&  0.3&  0.45&  0.07&  0.13&  0.09\\
14& 10.0&  0.3&  0.22&  0.04&  0.36&  0.06\\
15&  9.0&  0.5&  0.26&  0.04&  0.12&  0.05\\
16&  9.8&  0.2&  0.33&  0.08&  0.13&  0.09\\
17&  9.3&  0.2& -0.15&  0.05&  0.09&  0.08\\
18&  9.7&  0.3&  0.24&  0.08&  0.30&  0.06\\
19&  9.9&  0.3&  0.18&  0.03&  0.17&  0.04\\
20&  9.8&  0.3& -0.12&  0.09&  0.18&  0.14\\
21& $\gg$15&- &     -&     -&     -&     -\\
22&   3$<$&  -&    -1&     -&     -&     -\\
23& 10.2&  0.3&  0.22&  0.04&  0.30&  0.06\\
24&  9.6&  0.3&  0.53&  0.07&  0.04&  0.07\\
25&  3$<$&   -&     -&     -&     -&   -  \\
26&  $\gg$15&-&     -&     -&     -&     -\\

\enddata
\end{deluxetable}

\clearpage
\begin{figure}[bp]
\epsscale{1.1}
\mbox{\plottwo{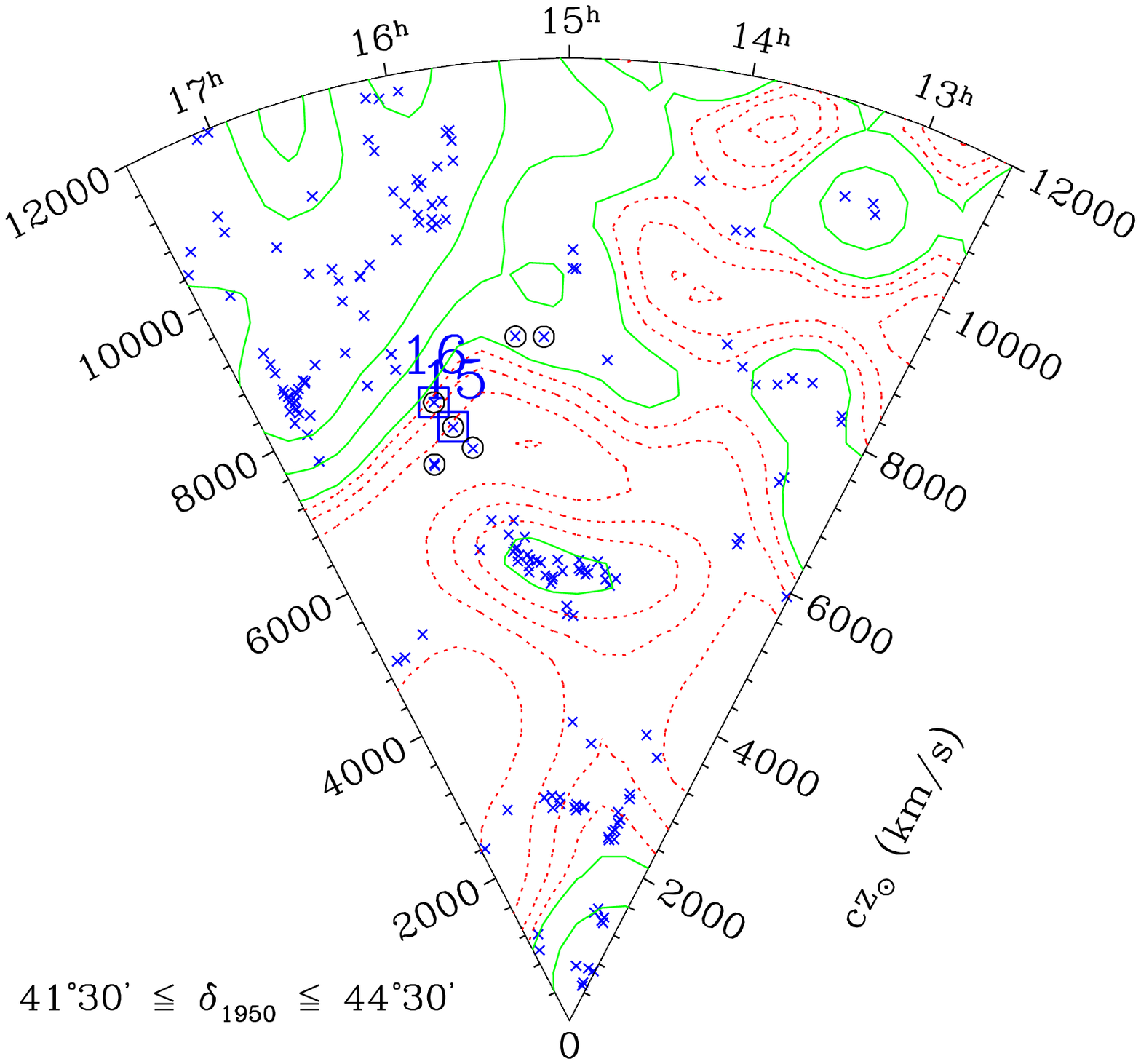}{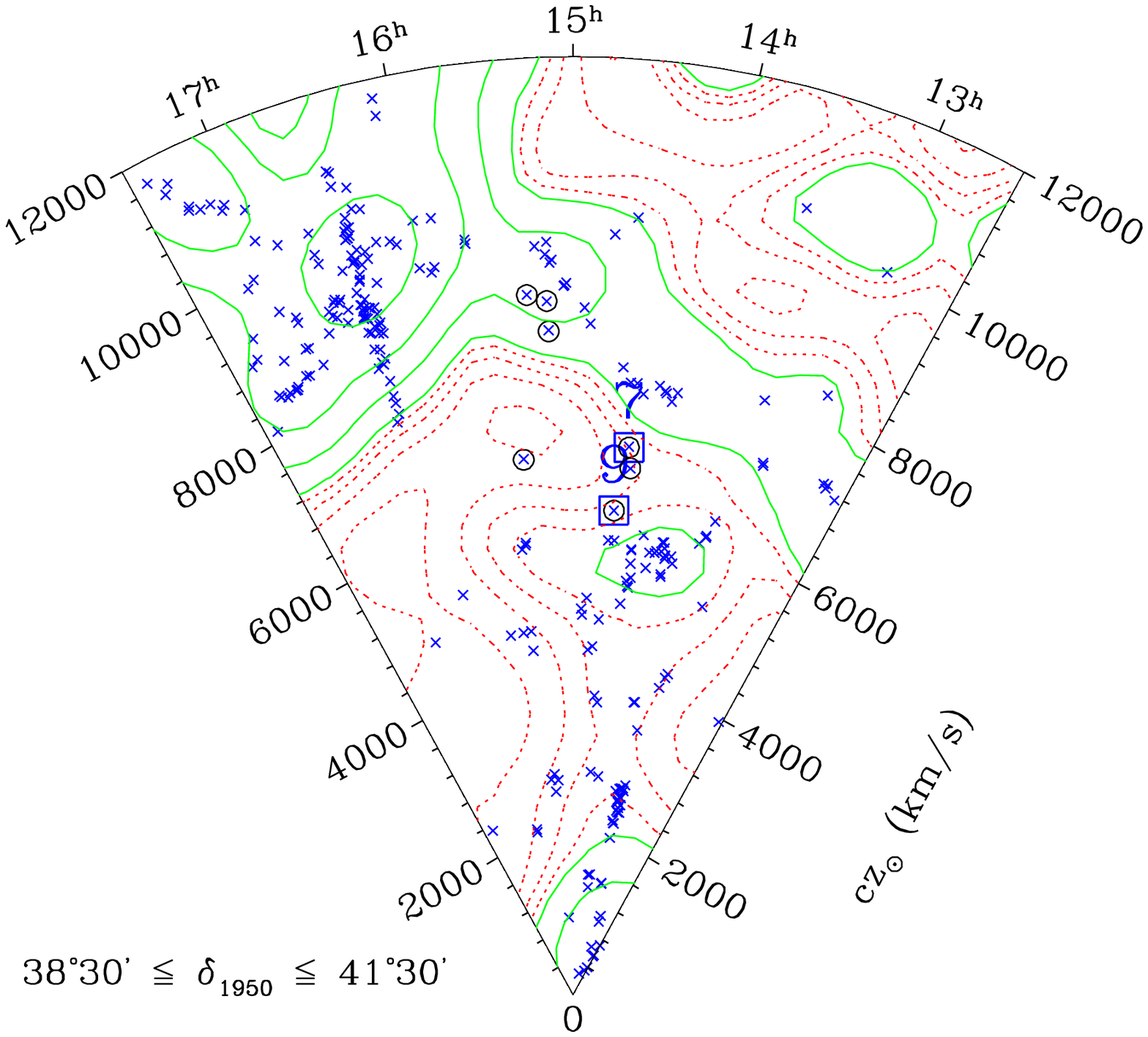}}
\mbox{\plottwo{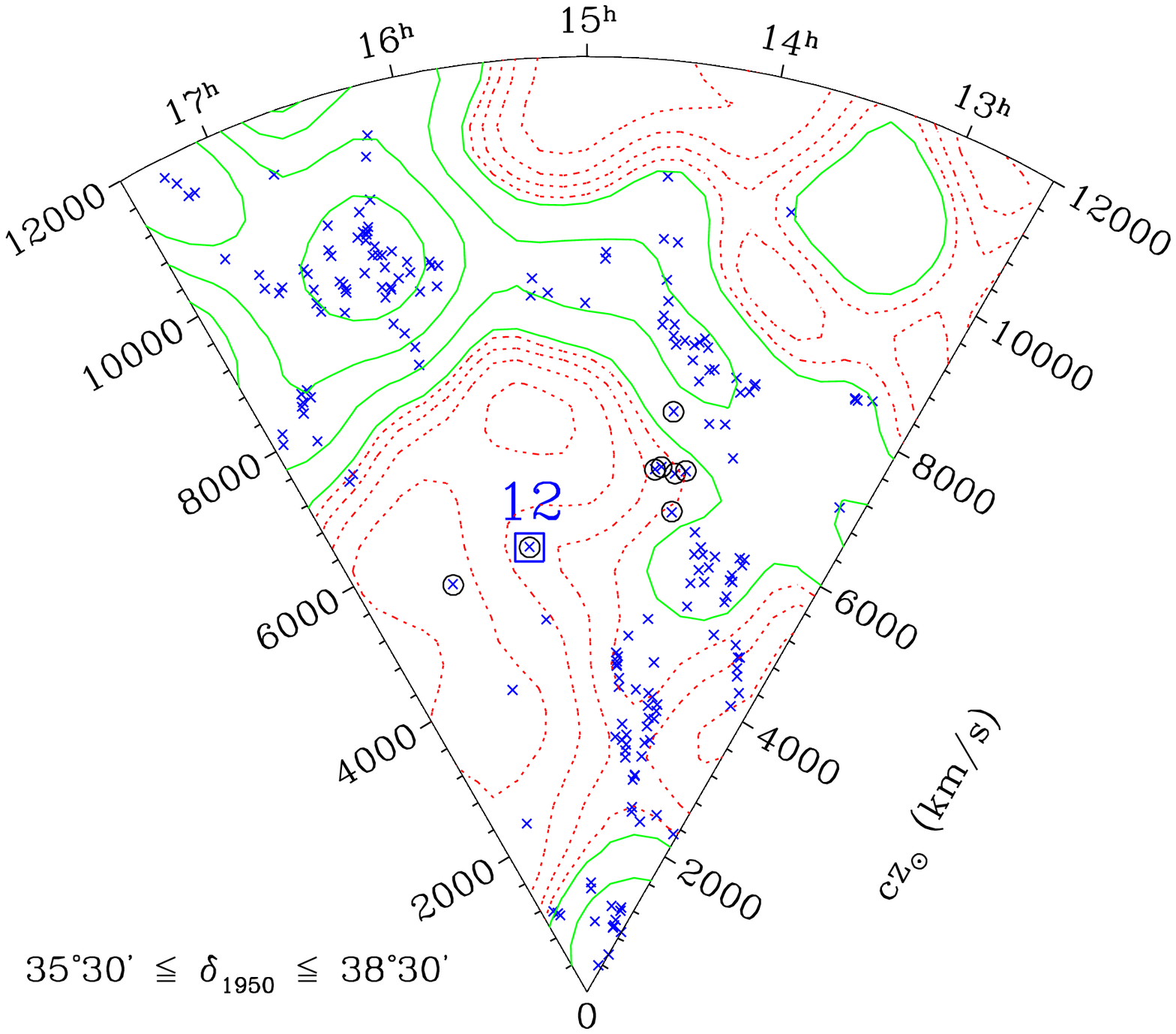}{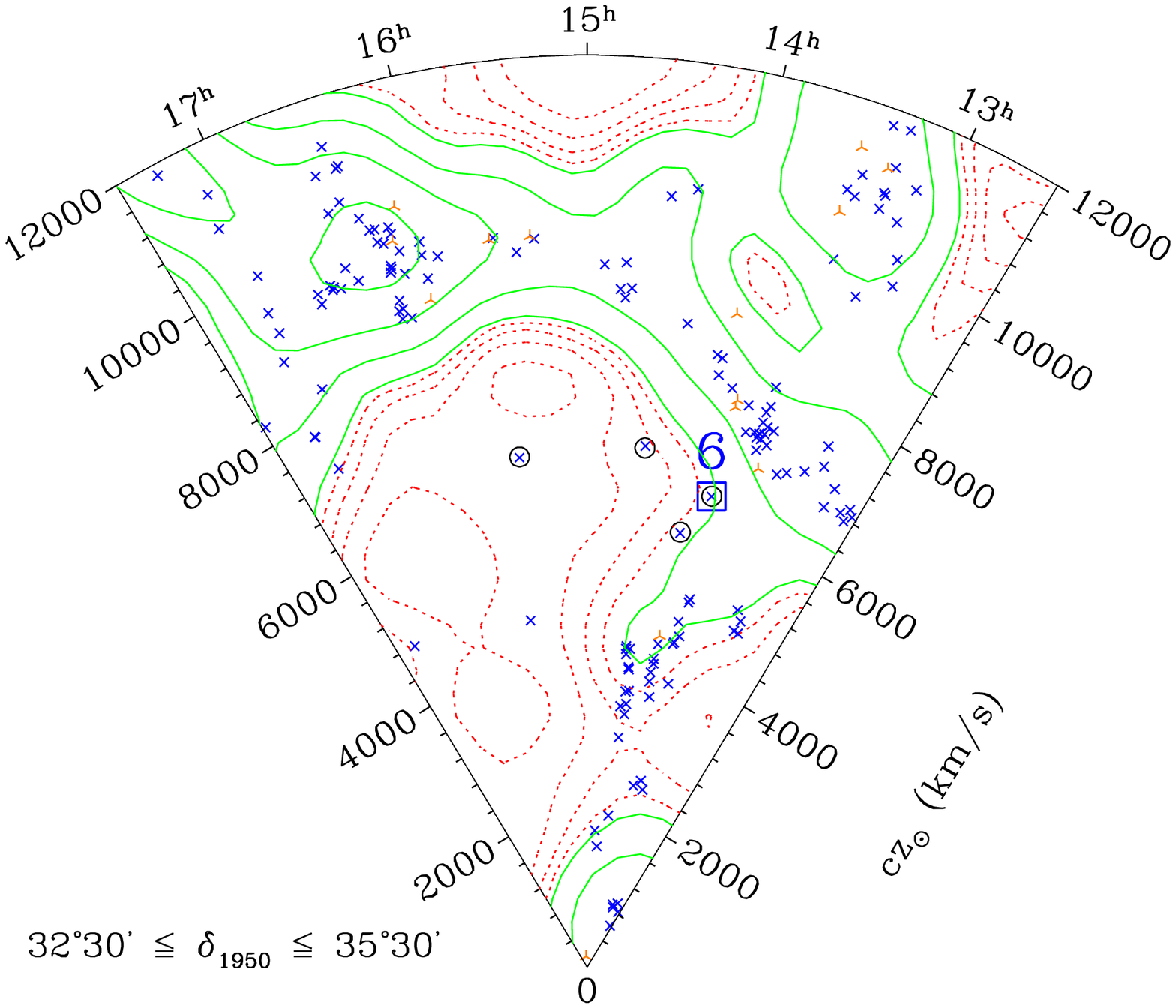}}
\caption{\footnotesize Eight successive $3\arcdeg$ declination slices
delineating the northern void in this study.  The CfA2 redshift survey 
galaxies are 
plotted with crosses, fainter 15R redshift survey galaxies with 
orange triangular crosses.
The GG99 sample is circled.  The subset of GG99 investigated in this study is 
marked 
by larger squares with labels above, referring to the reference number in 
Table~1\@. Blue and orange numbers are from CfA2 and 15R respectively.
We overplot $5h^{-1}$ Mpc-smoothed number
density contours as determined from CfA2.  Underdensities in $0.2\bar
n$ decrements are marked with dotted contours; overdensities in
logarithmic intervals of $\bar n$, $2\bar n$, $4\bar n$, etc., are
marked with solid contours. 
}
\label{cfa2nfig}
\end{figure}

\clearpage
\begin{figure}
\figurenum{\ref{cfa2nfig}}
\caption{\footnotesize Cont'd.}
\epsscale{1.15}
\mbox{\plottwo{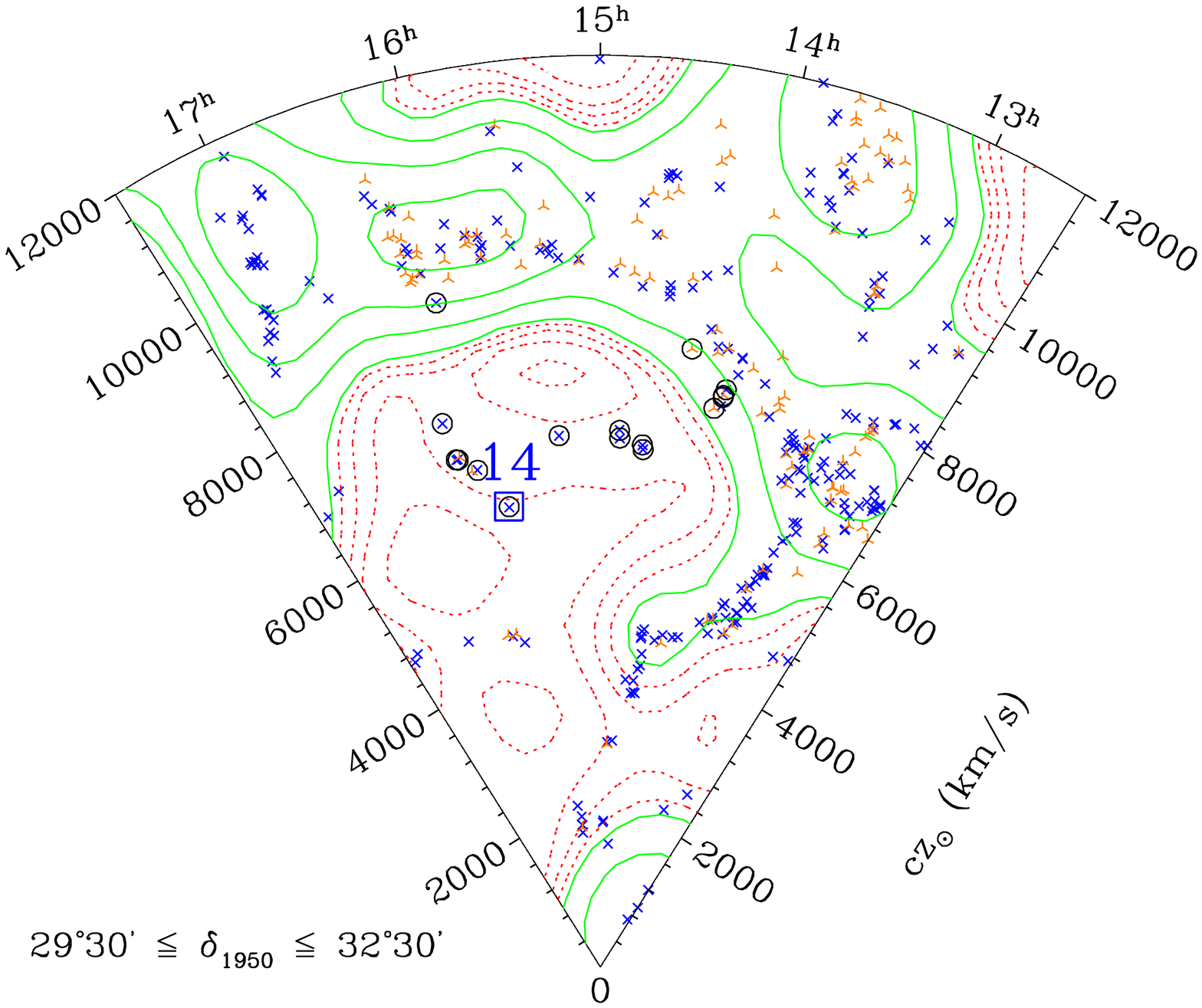}{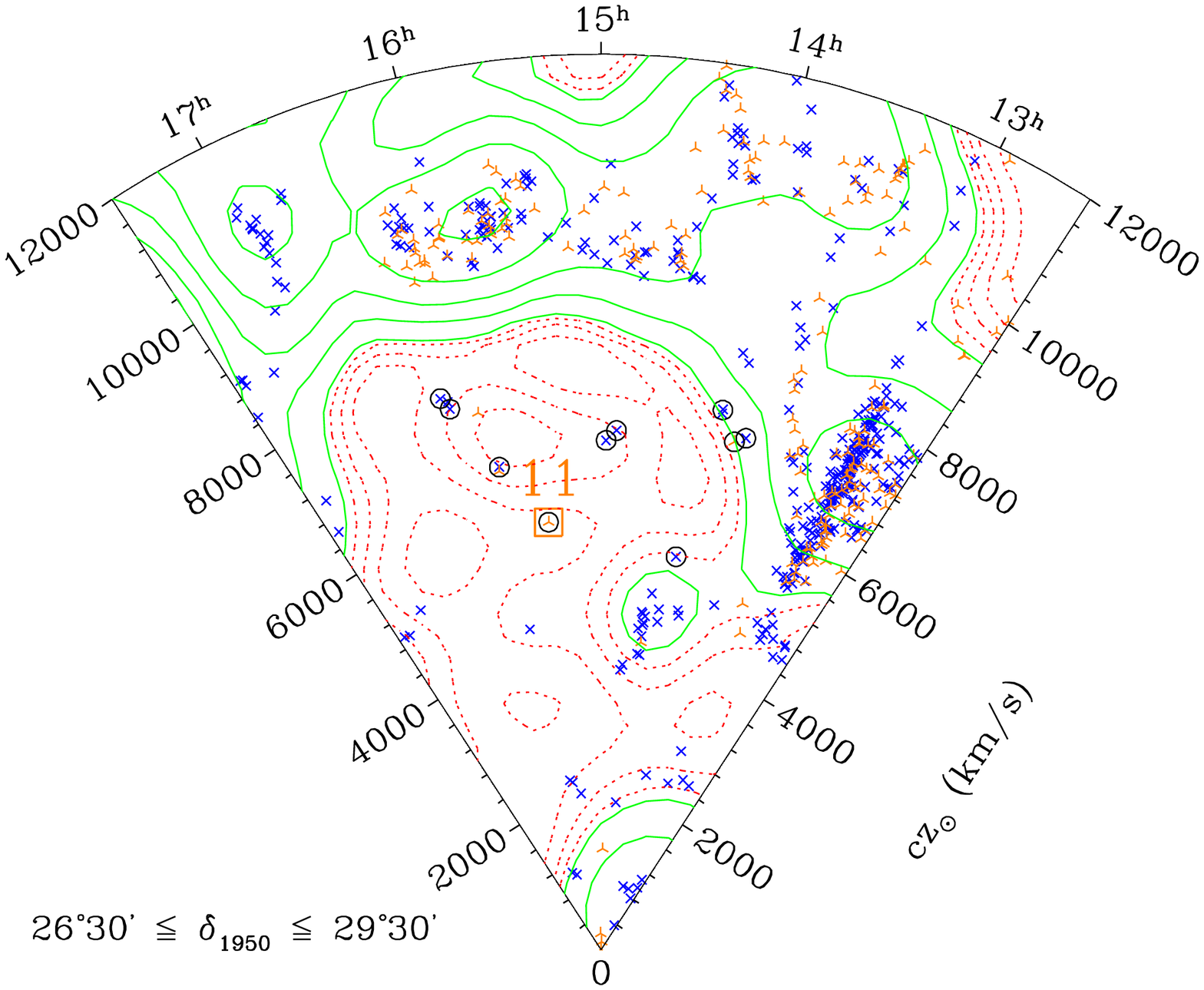}}
\mbox{\plottwo{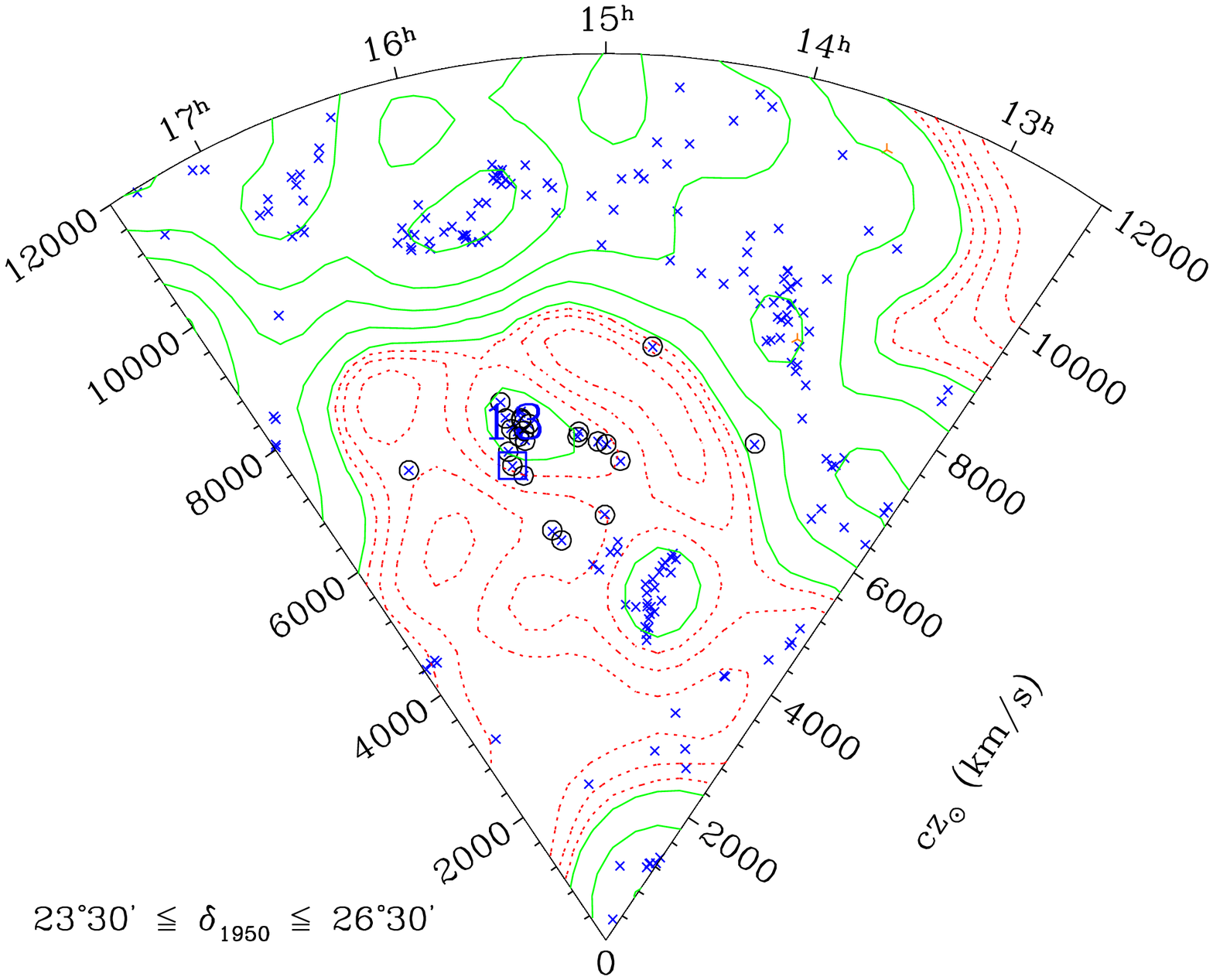}{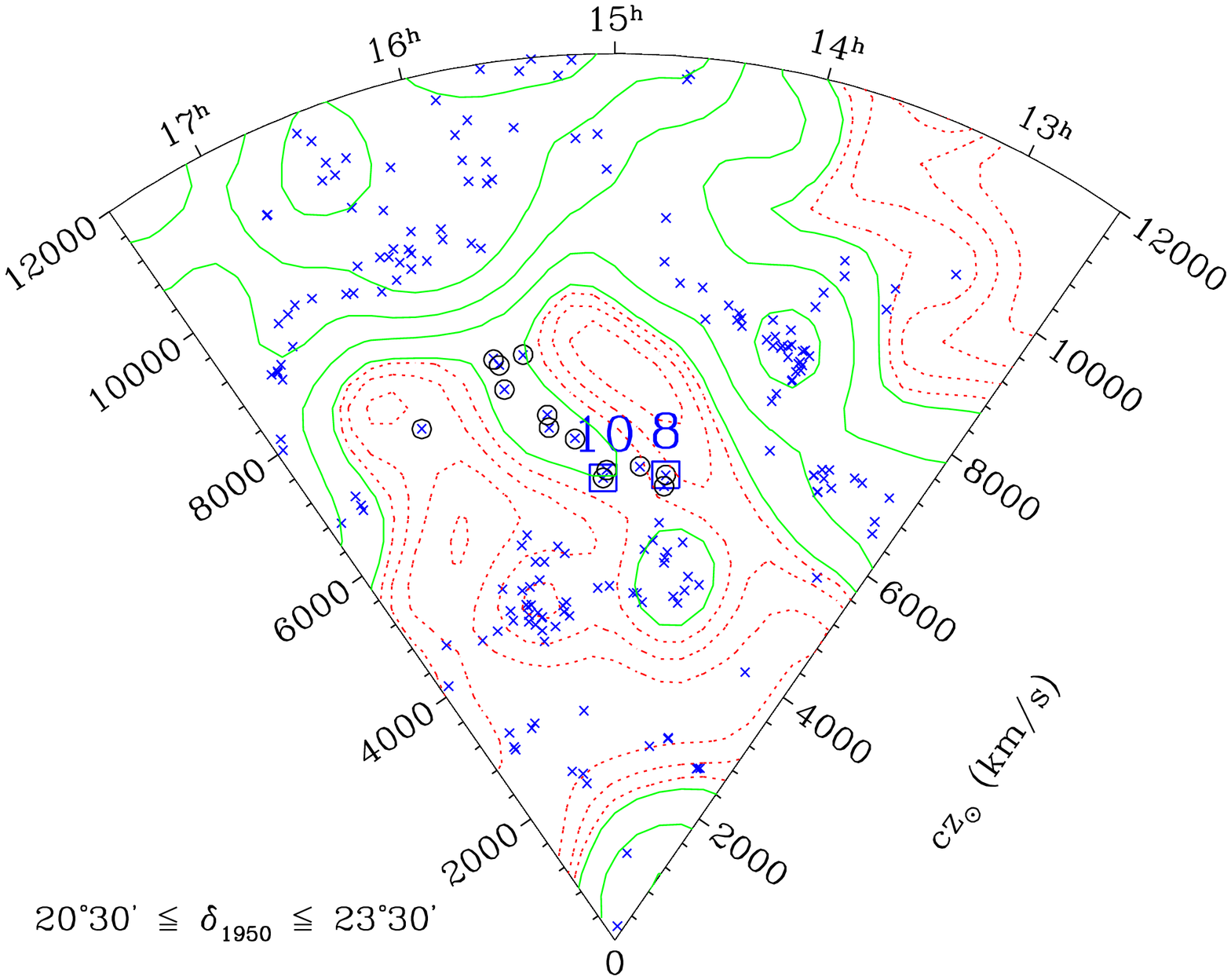}}
\end{figure}

\clearpage
\begin{figure}[bp]
\caption{\footnotesize As Fig.~1, for CfA2 South voids.
\label{cfa2sfig}}
\mbox{\plottwo{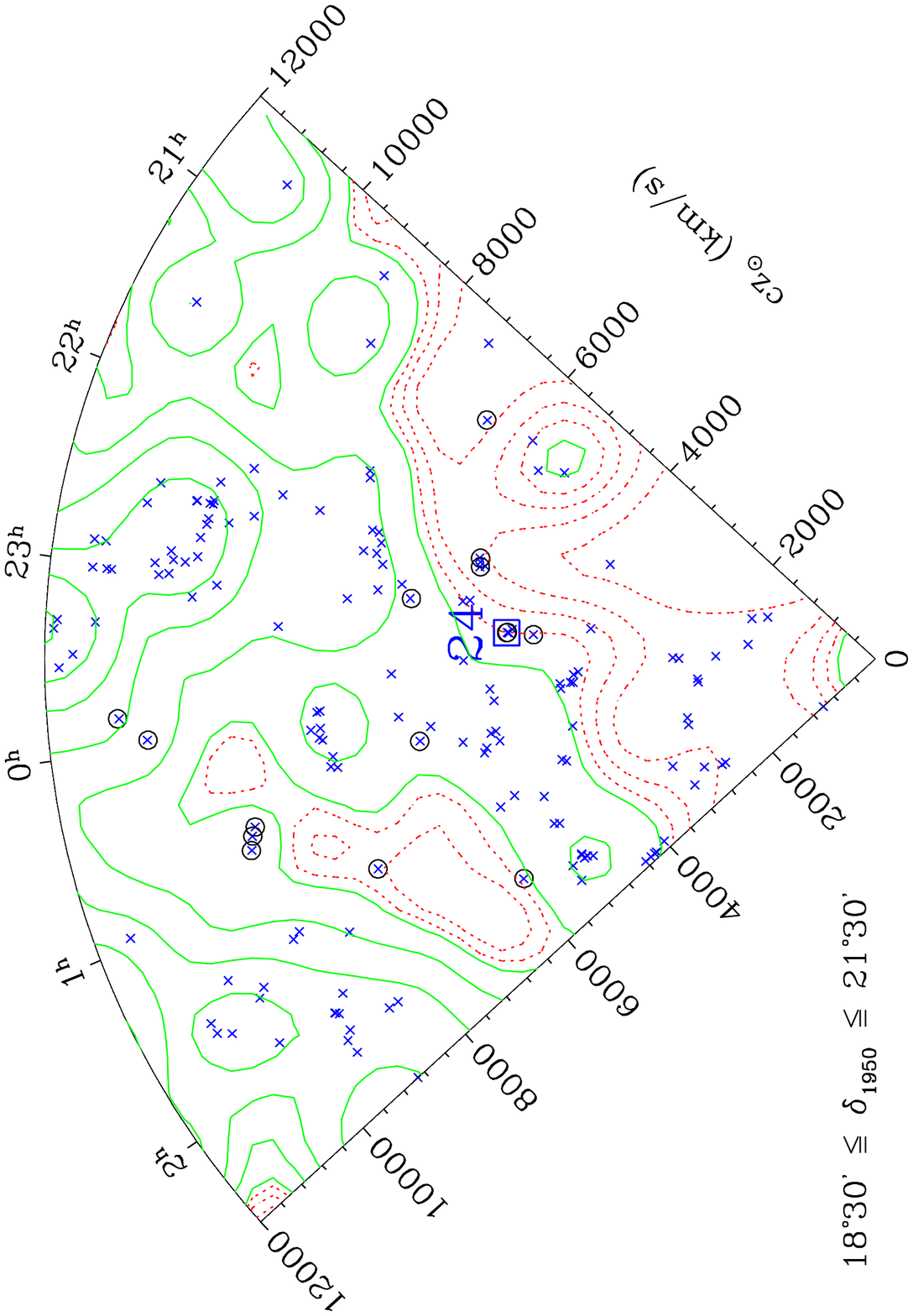}{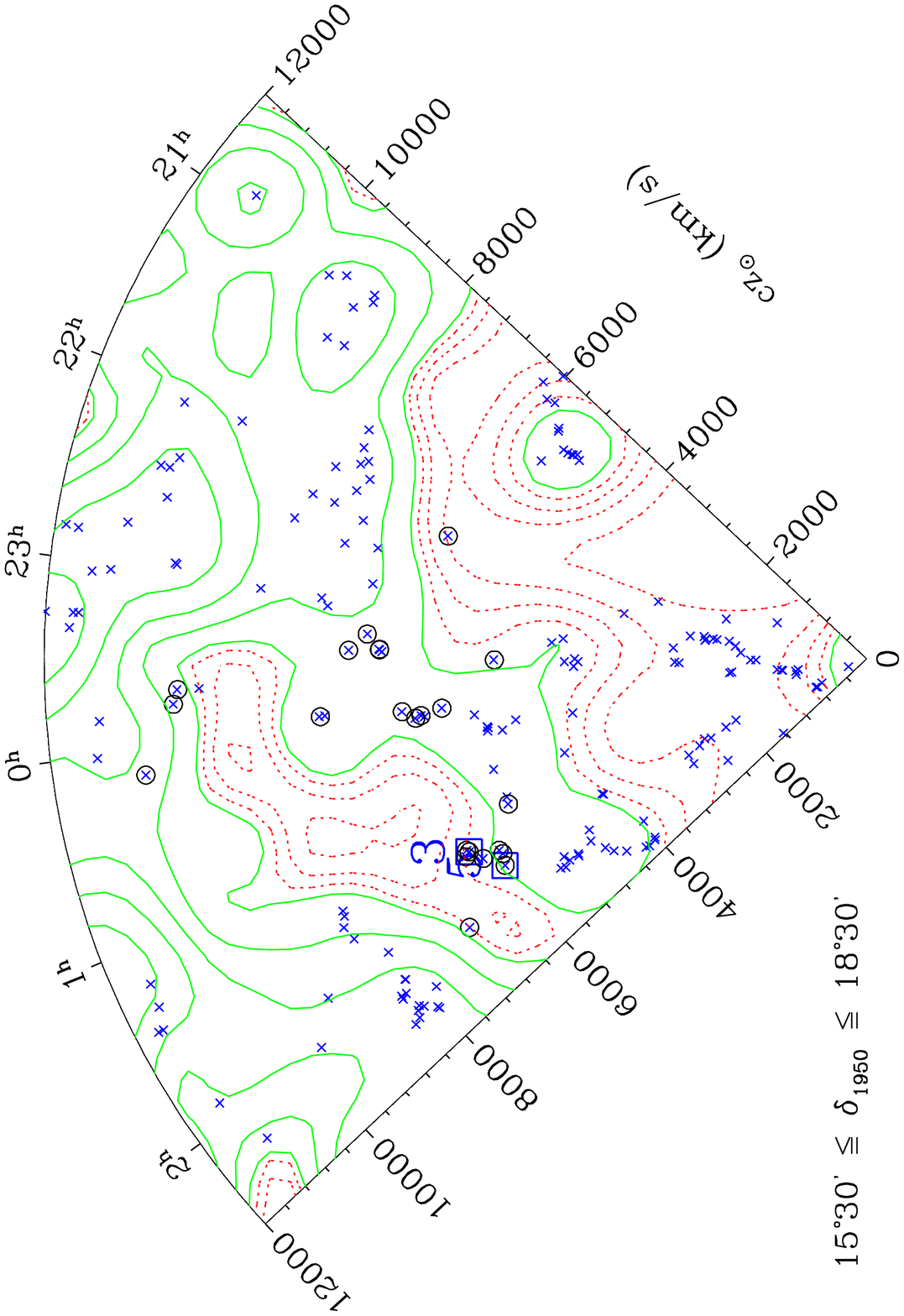}}
\mbox{\plottwo{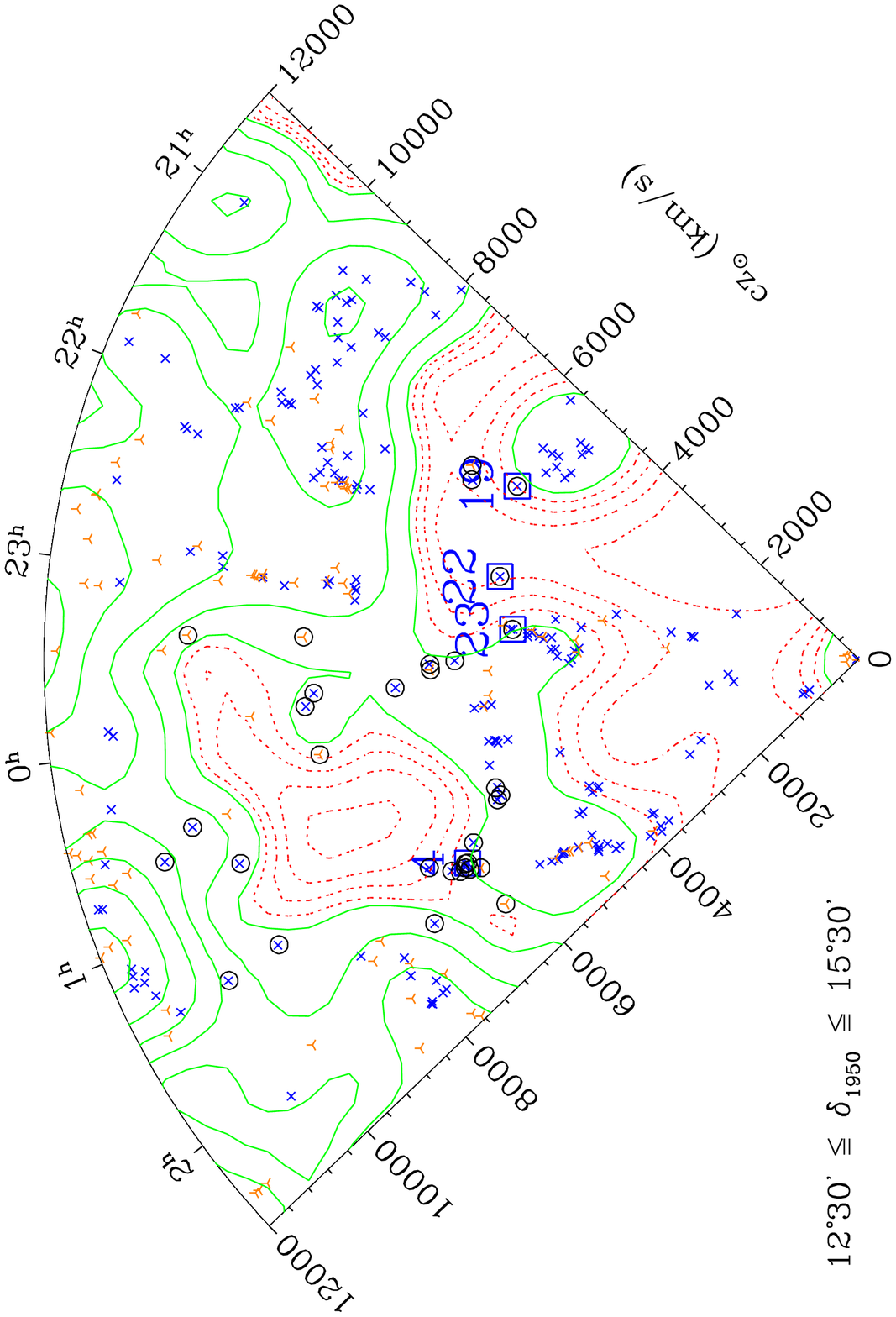}{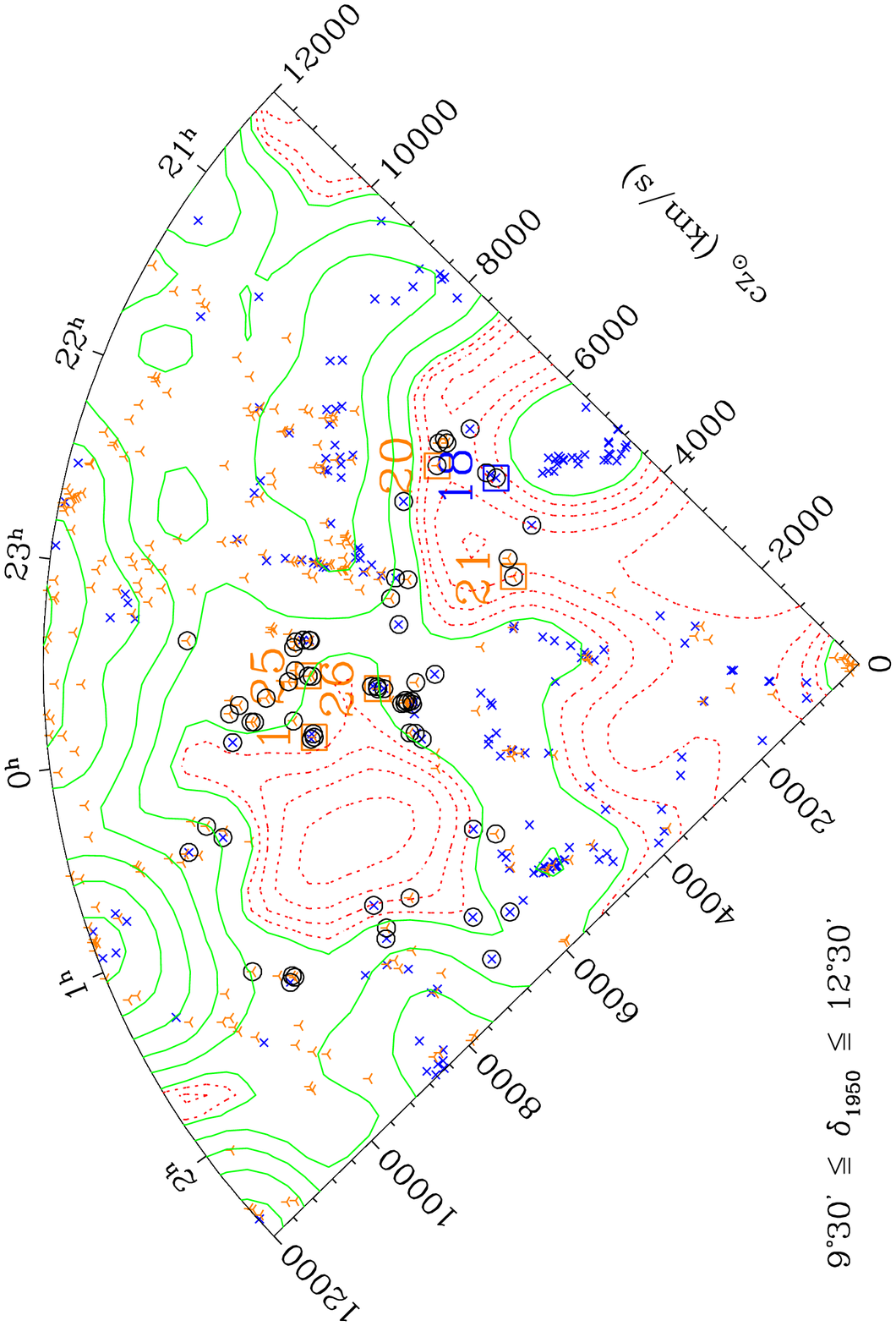} }
\end{figure}

\clearpage
\begin{figure}
\figurenum{\ref{cfa2sfig}}
\caption{\footnotesize Cont'd.}
\mbox{\plottwo{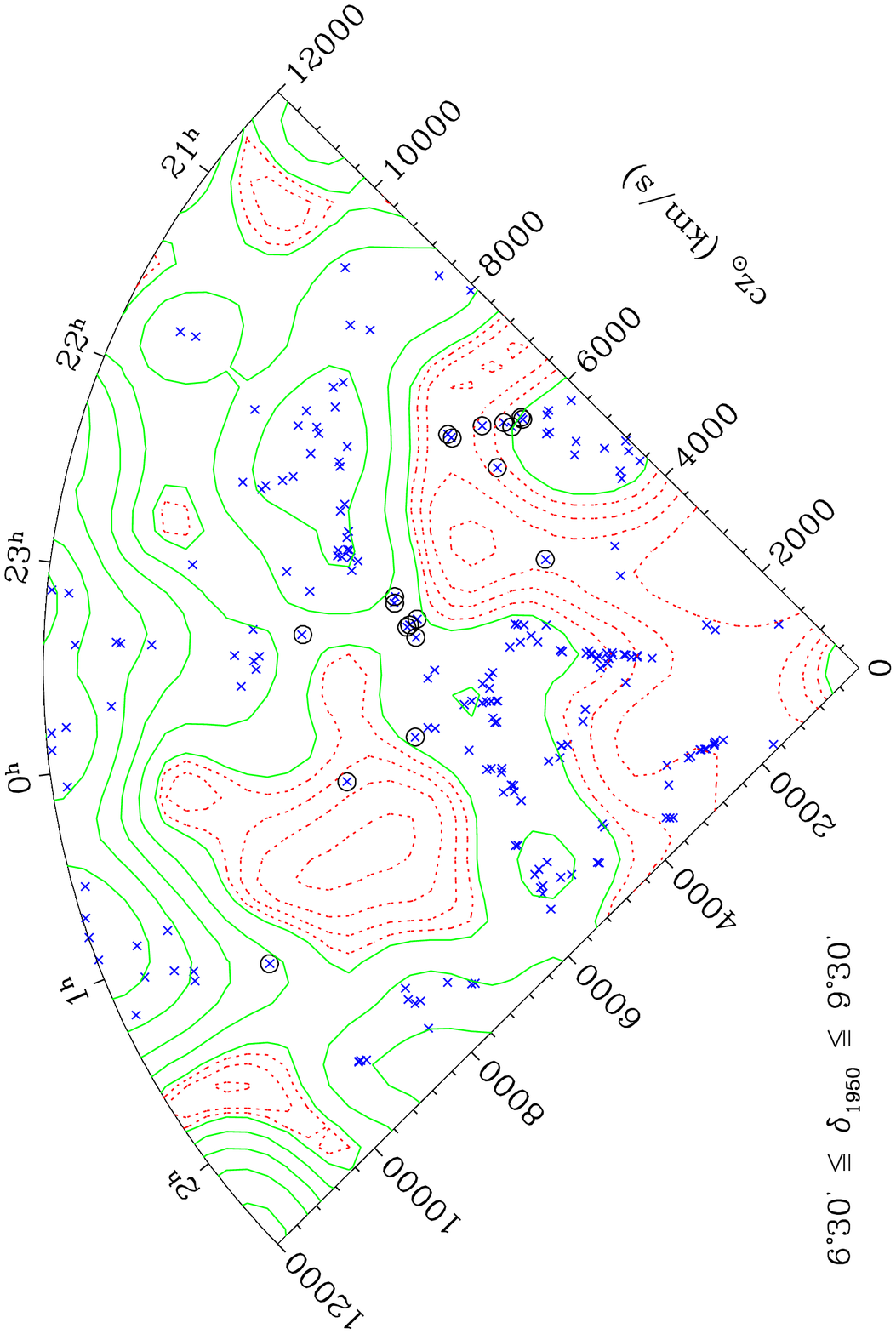}{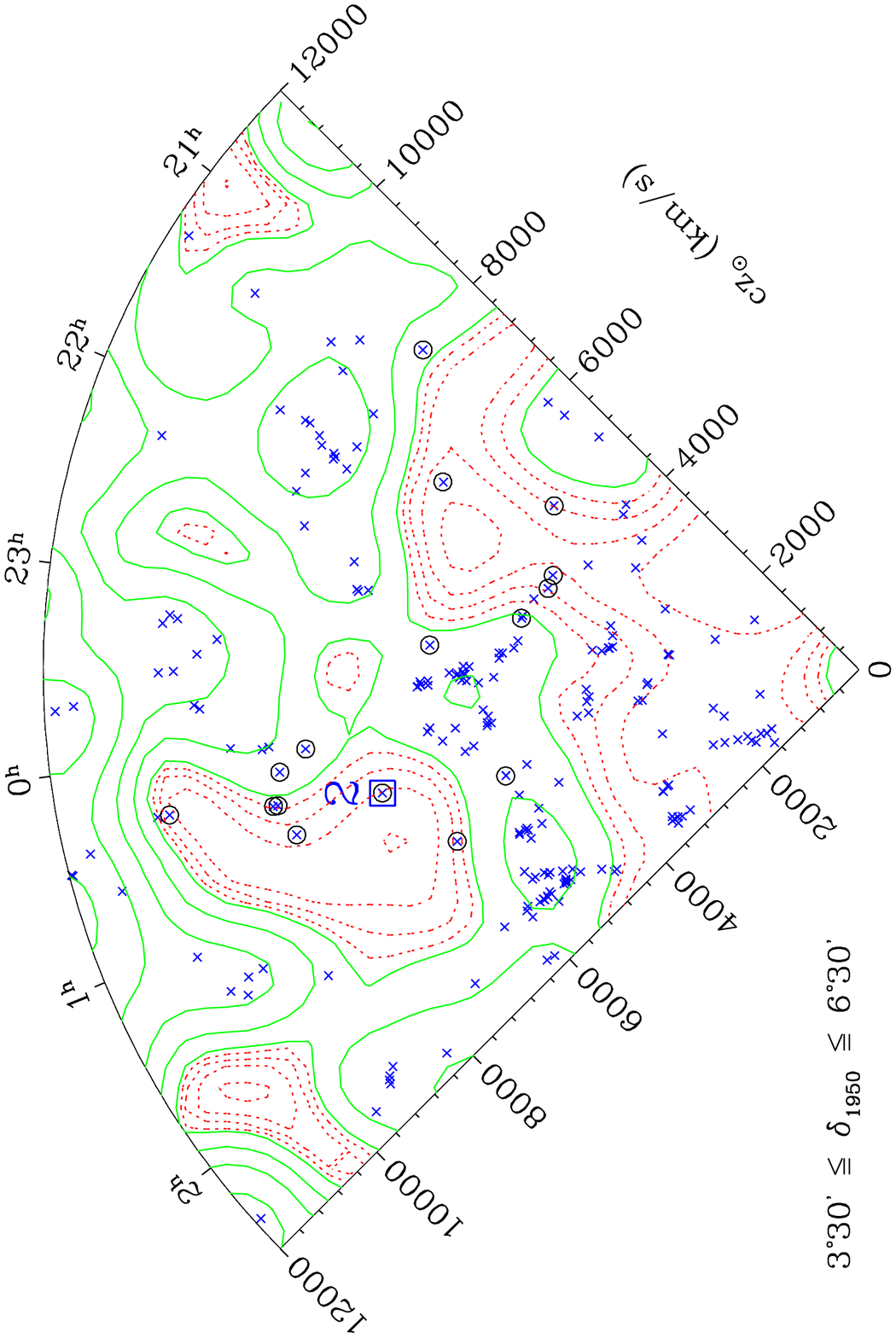} }
\mbox{\plottwo{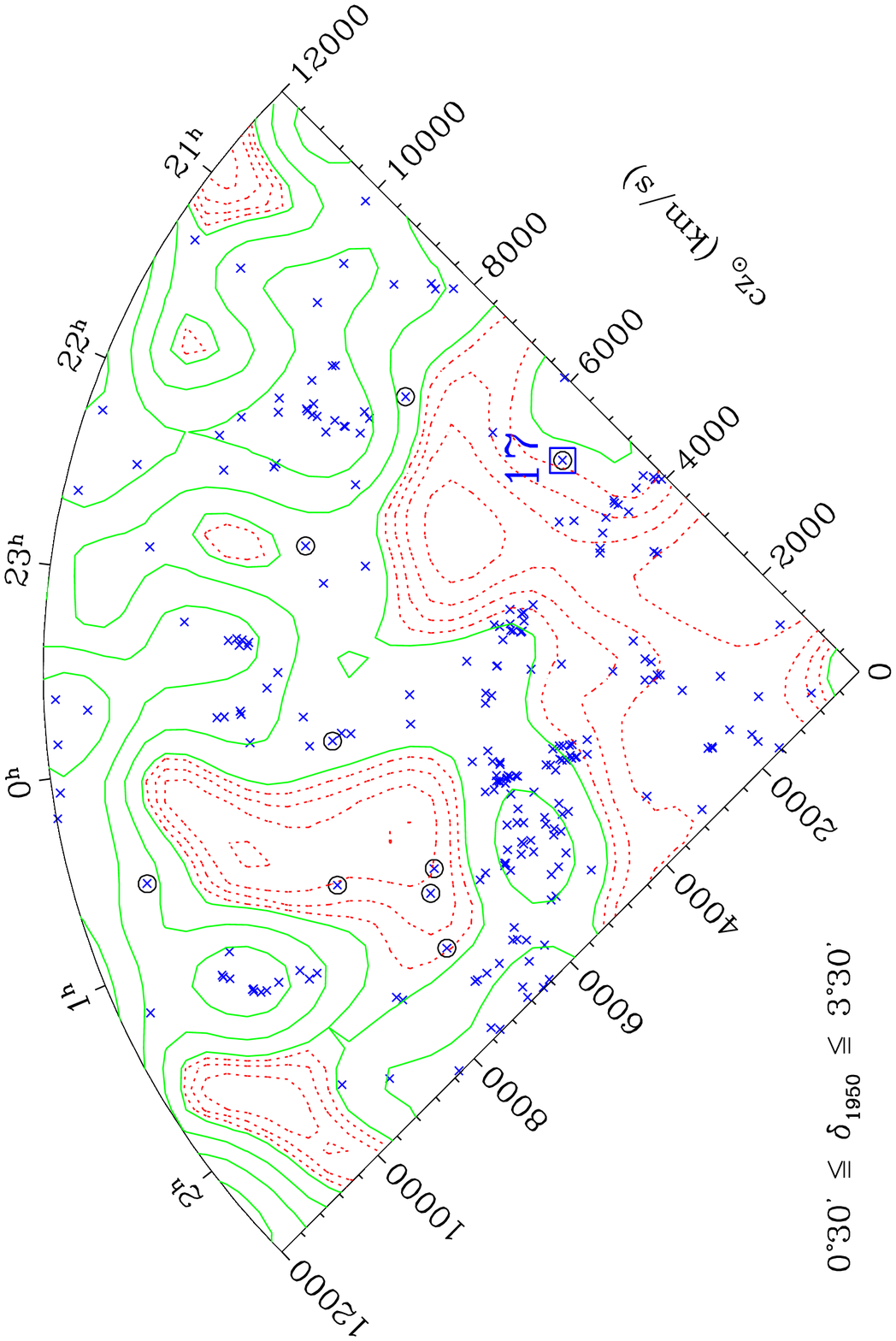}{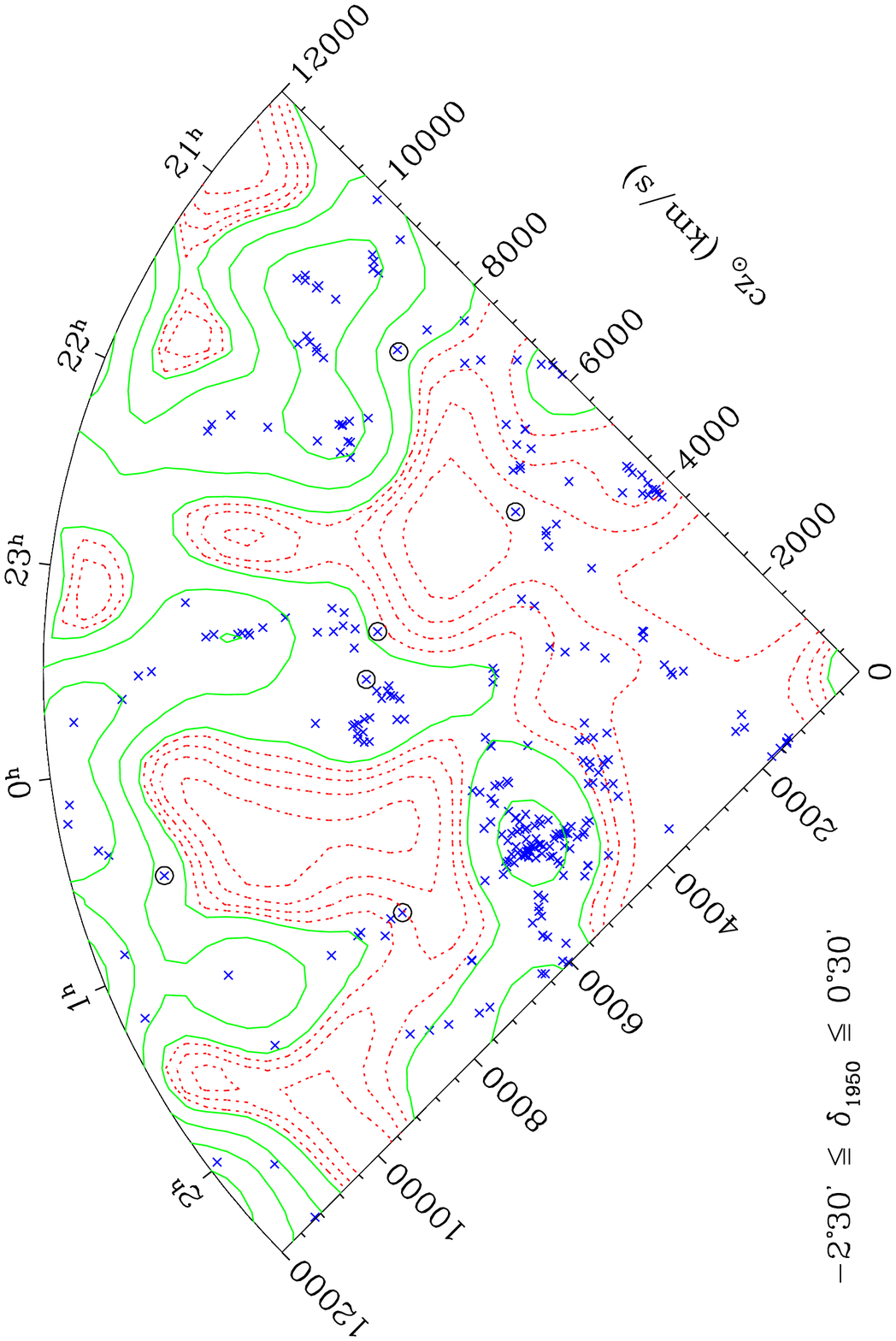}}
\end{figure}

\clearpage
\begin{figure}
\centering
\includegraphics[height=18truecm]{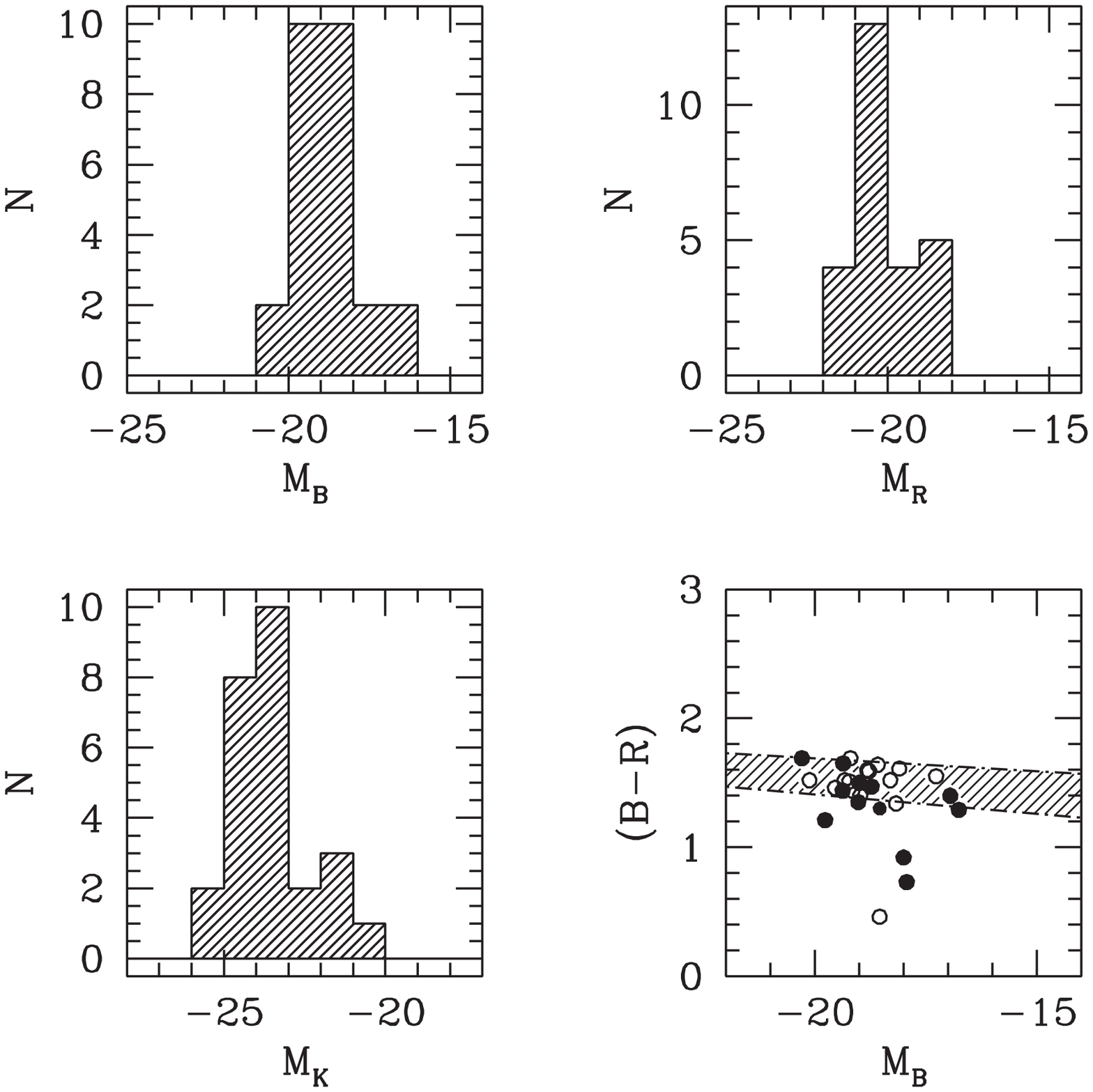}
\caption{\footnotesize Histograms of the total $B, R$, and $K$
magnitudes of the void galaxies and the color magnitude diagram.
Objects showing H$\beta$ emission are shown as filled circles and the
location of the red sequence in the Coma cluster estimated from
Mobasher et al. (2001)
is indicated. 
}

\end{figure}

\clearpage
\begin{figure} [bp]
\label{pospix}
\includegraphics[height=20truecm]{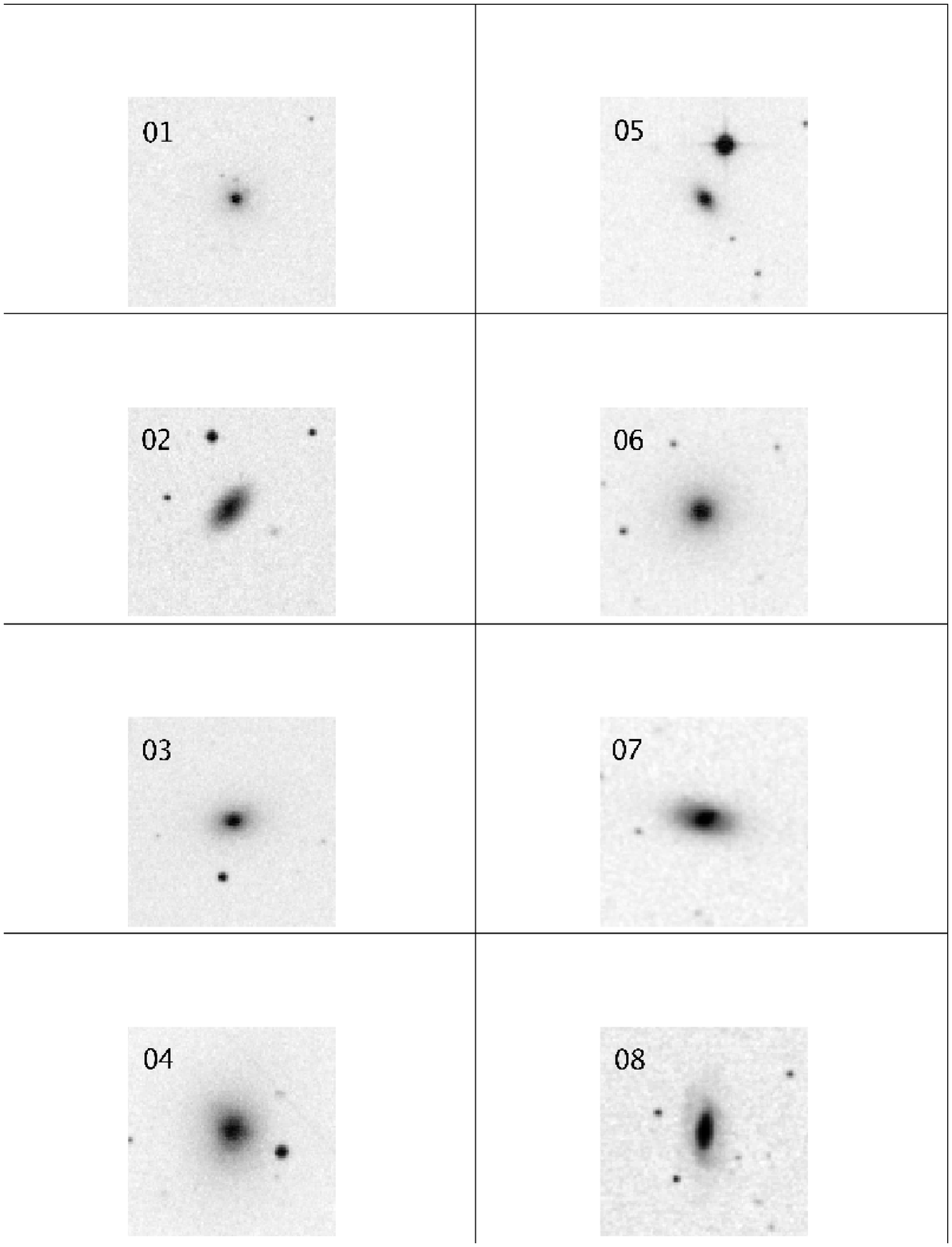}
\caption{\footnotesize Images of the VGs from the digitized POSS2 blue
data as referenced in the text. North is to the top and East is to the left.
Each picture is $2^\prime~ X~ 2^\prime$ on a side.
The compressed files of the ``Palomar Observatory - Space Telescope Science
Institute Digital Survey'' of the northern sky, based on scans of the 
Second Palomar Sky Survey are copyright $\copyright$ 1993-1995 by the
California Institute of Technology.
}
\end{figure}

\clearpage
\begin{figure}
\figurenum{4}
\includegraphics[height=20truecm]{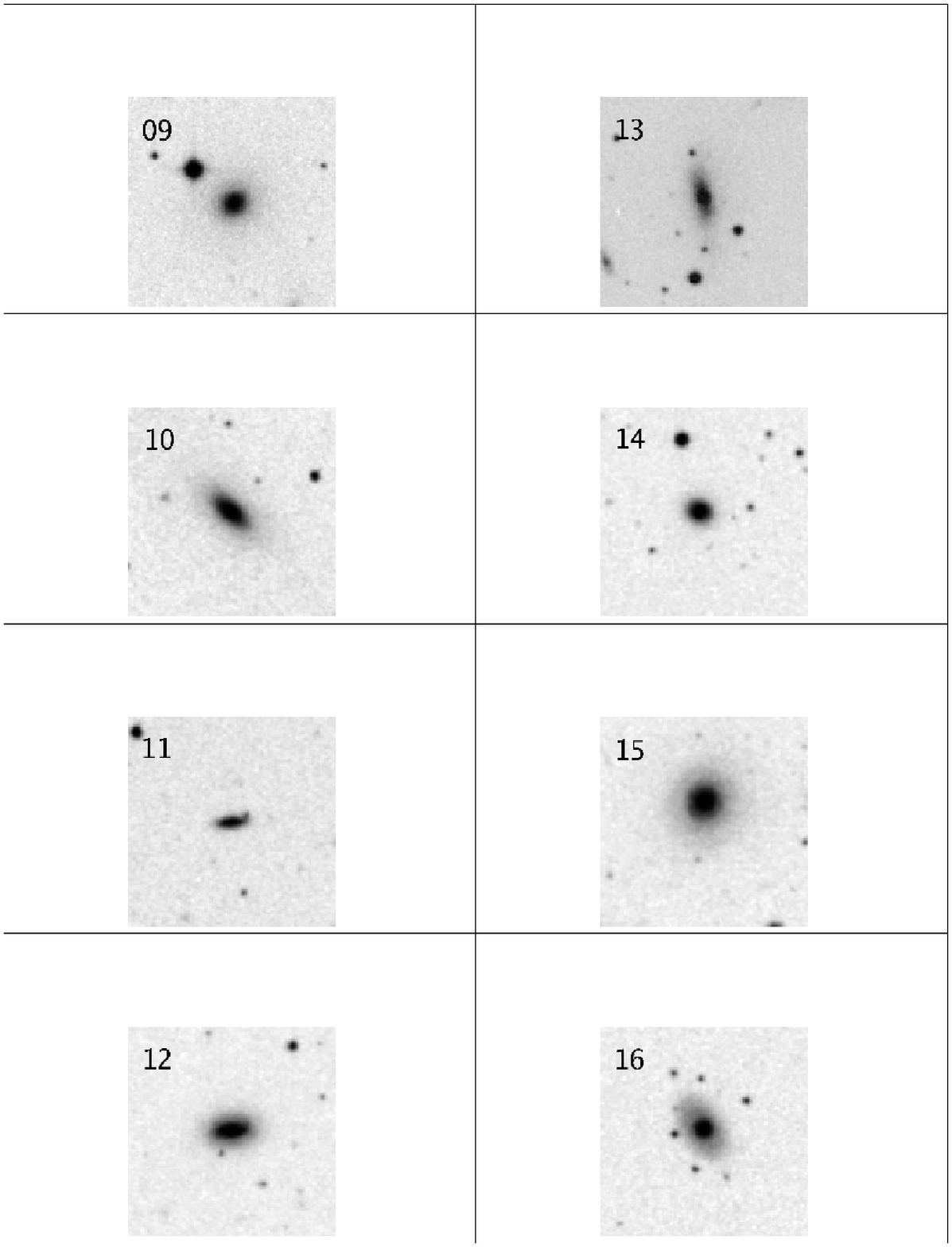}
\caption{\footnotesize Cont'd. 
}
\end{figure}

\clearpage
\begin{figure}
\centering
\figurenum{4}
\includegraphics[height=20truecm]{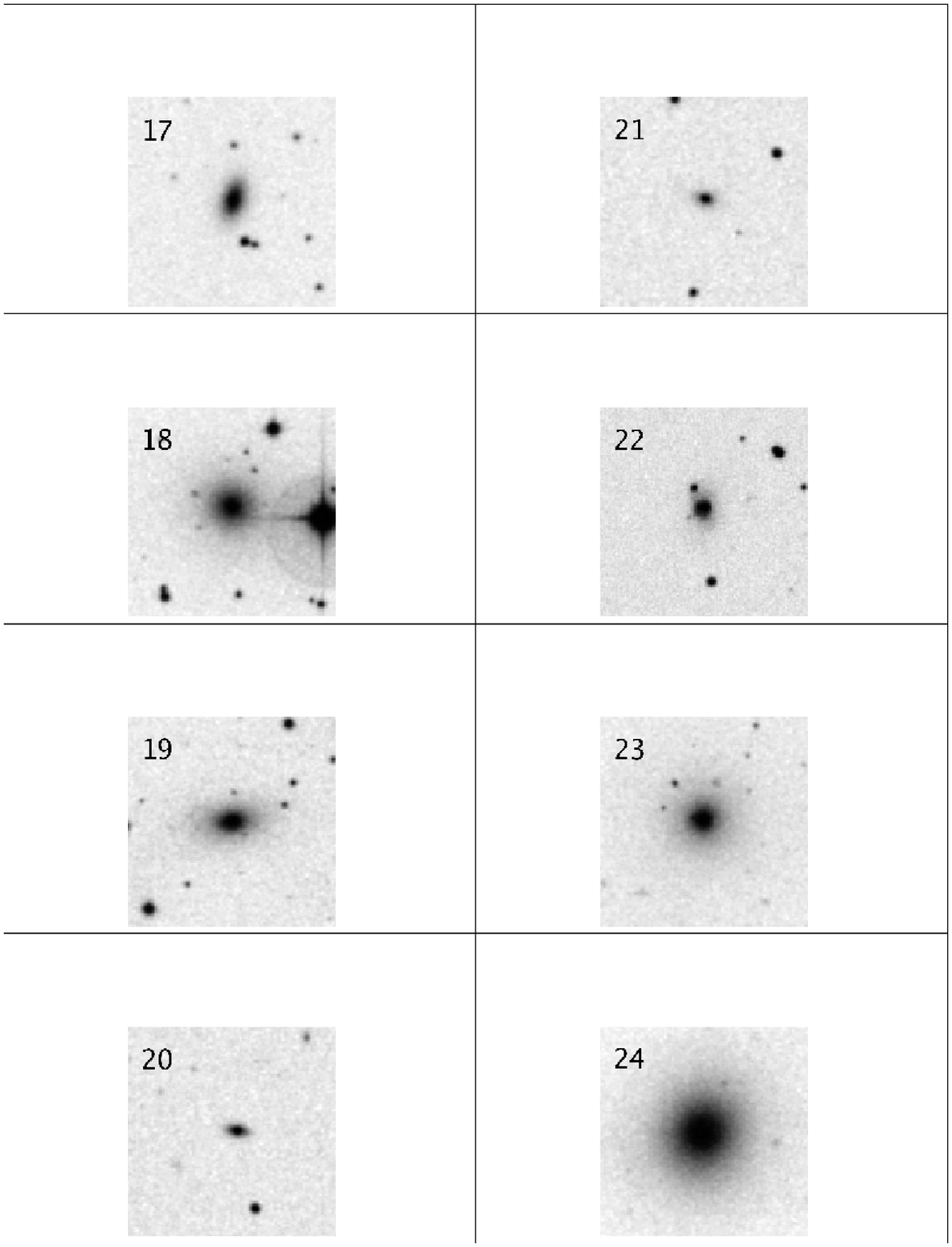}
\caption{\footnotesize Cont'd.
}
\end{figure}

\clearpage
\begin{figure}
\centering
\figurenum{4}
\includegraphics[height=20truecm]{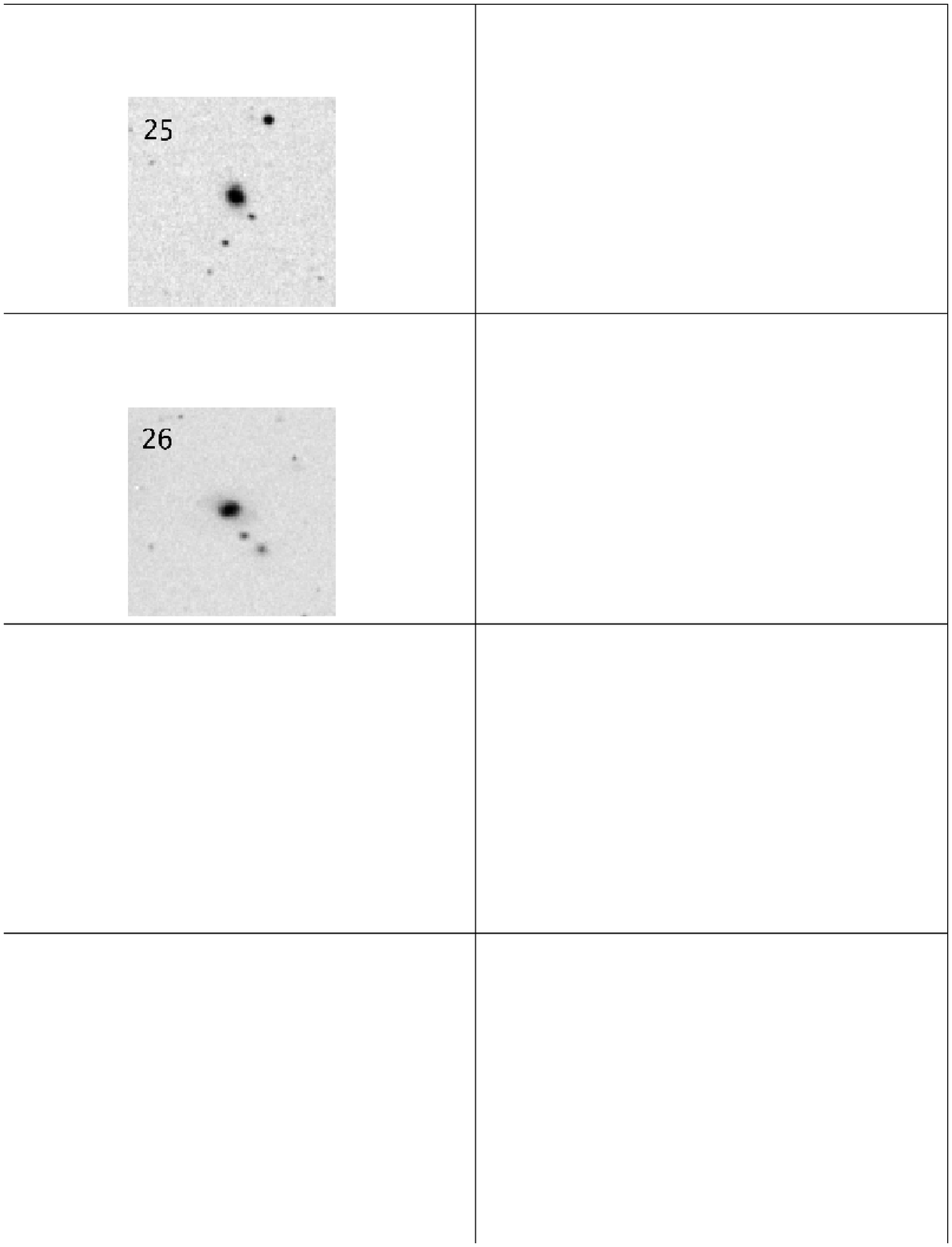}
\caption{\footnotesize Cont'd.
}
\end{figure}


\clearpage
\begin{figure}
\label{speks}
\includegraphics[height=18truecm]{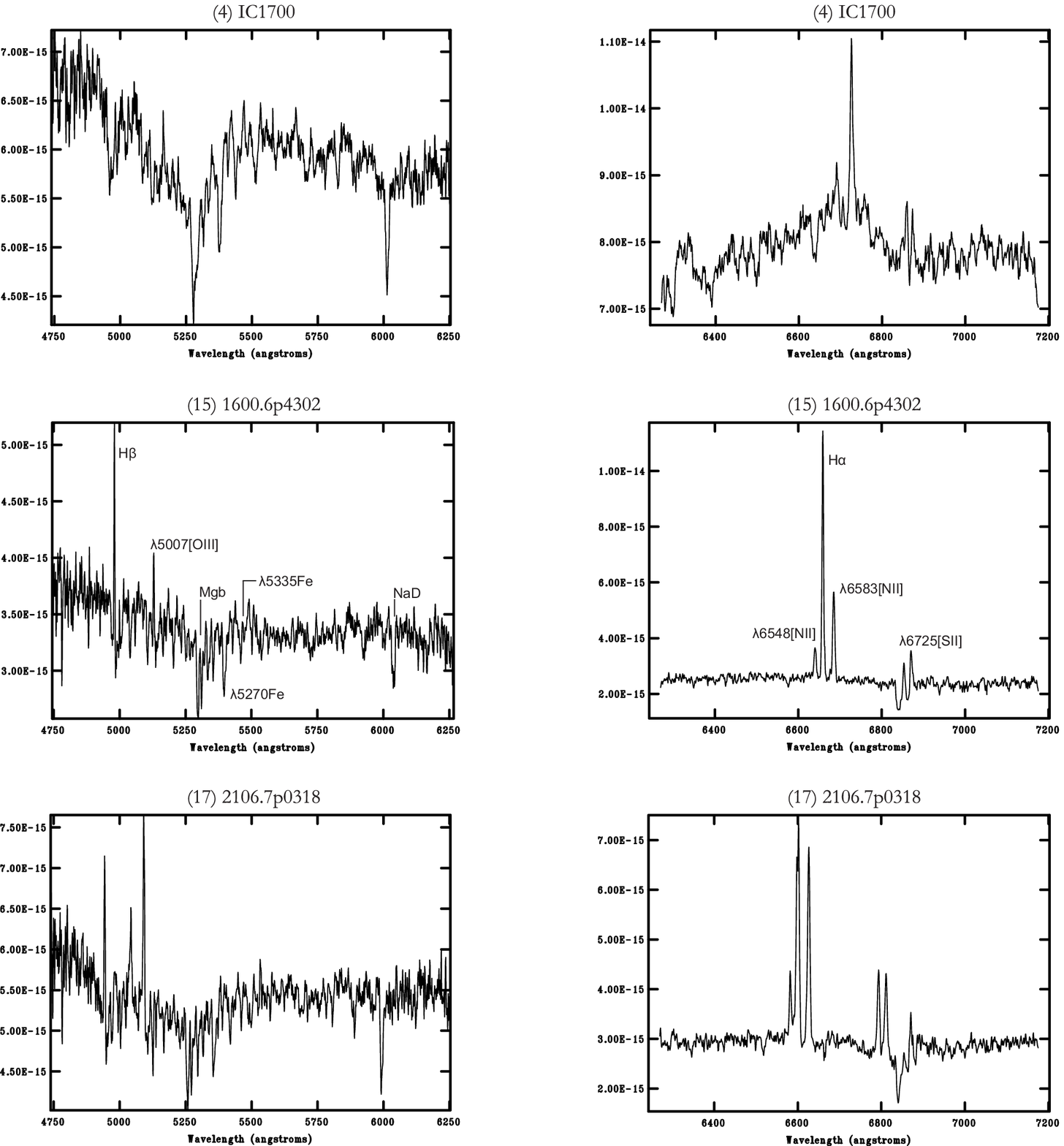}
\caption{\footnotesize Examples of some of the VG spectra. Blue
portion to the left and red to the right. The spectra have been flux
corrected, but flux values are relative as not all objects were 
observed under photometric conditions. Some of the more important
spectral features that occur in all of the spectra
are indicated in the middle panels for the
galaxy 1600.6p4302.
}
\end{figure}

\clearpage
\begin{figure}
\figurenum{5}
\includegraphics[height=18truecm]{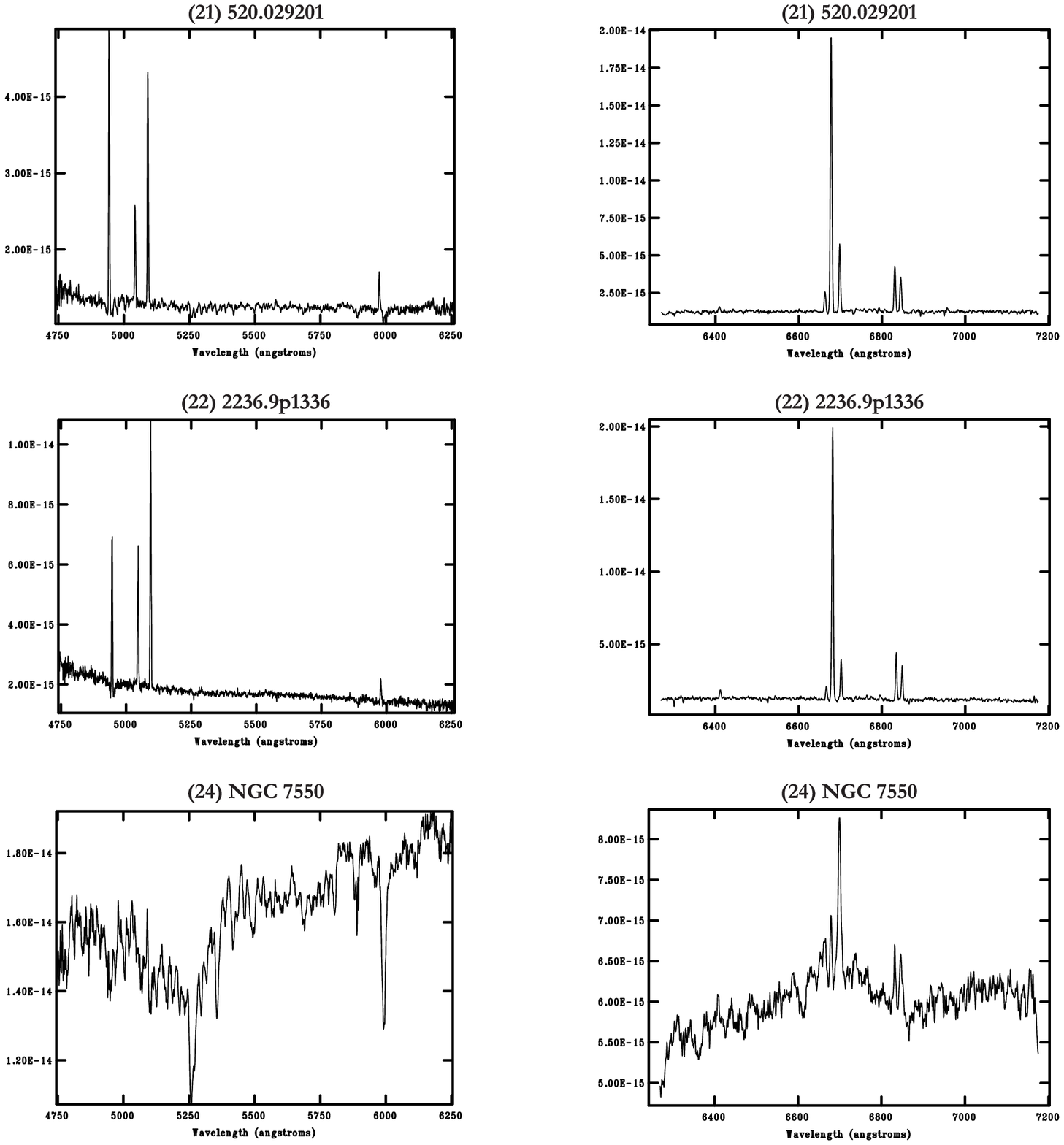}
\caption{\footnotesize Cont'd. 
}
\end{figure}


\clearpage
\begin{figure}
\centering
\includegraphics[height=12truecm]{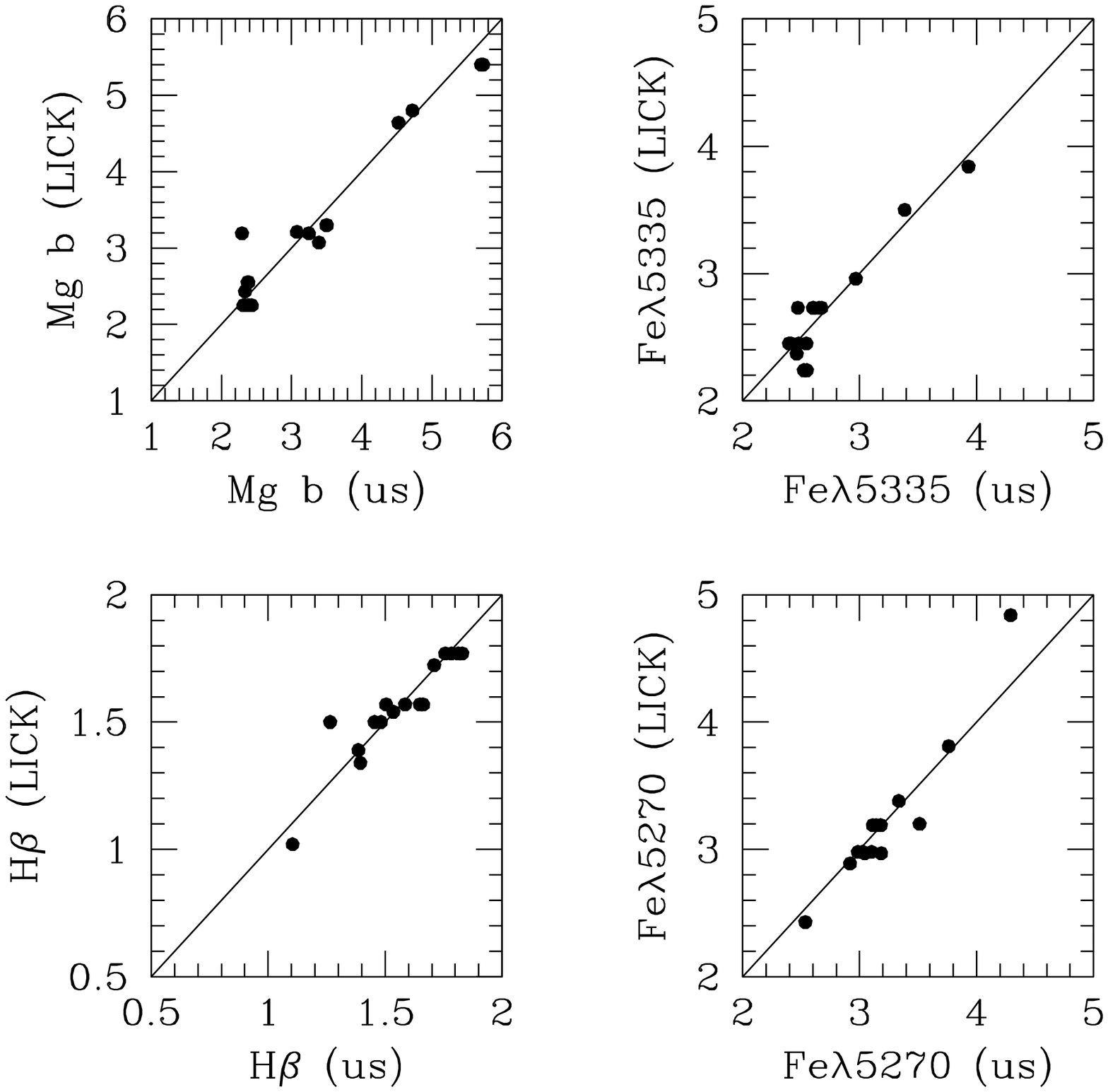}
\caption{\footnotesize Comparison between the line indices measured in 
this study with the Lick indices in Worthey \etal (1994)
and Trager \etal. (1998).
}
\end{figure}


\clearpage
\begin{figure}
\centering
\includegraphics[height=12truecm]{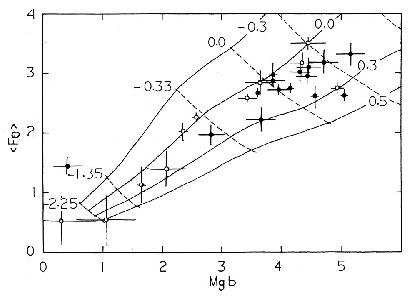}
\caption{\footnotesize
The positions of the observed galaxies with their error bars. The curves use
the TMB models for an assumed age of 12 Gyr. Solid curves denote loci of 
constant $[\alpha/Fe]$ and dashed curves indicate loci of fixed $[Z/H]$.
Objects with detected H$\beta$ emission are open circles.
}
\label{mgbfe}
\end{figure}

\clearpage
\begin{figure}
\centering
\includegraphics[height=12truecm]{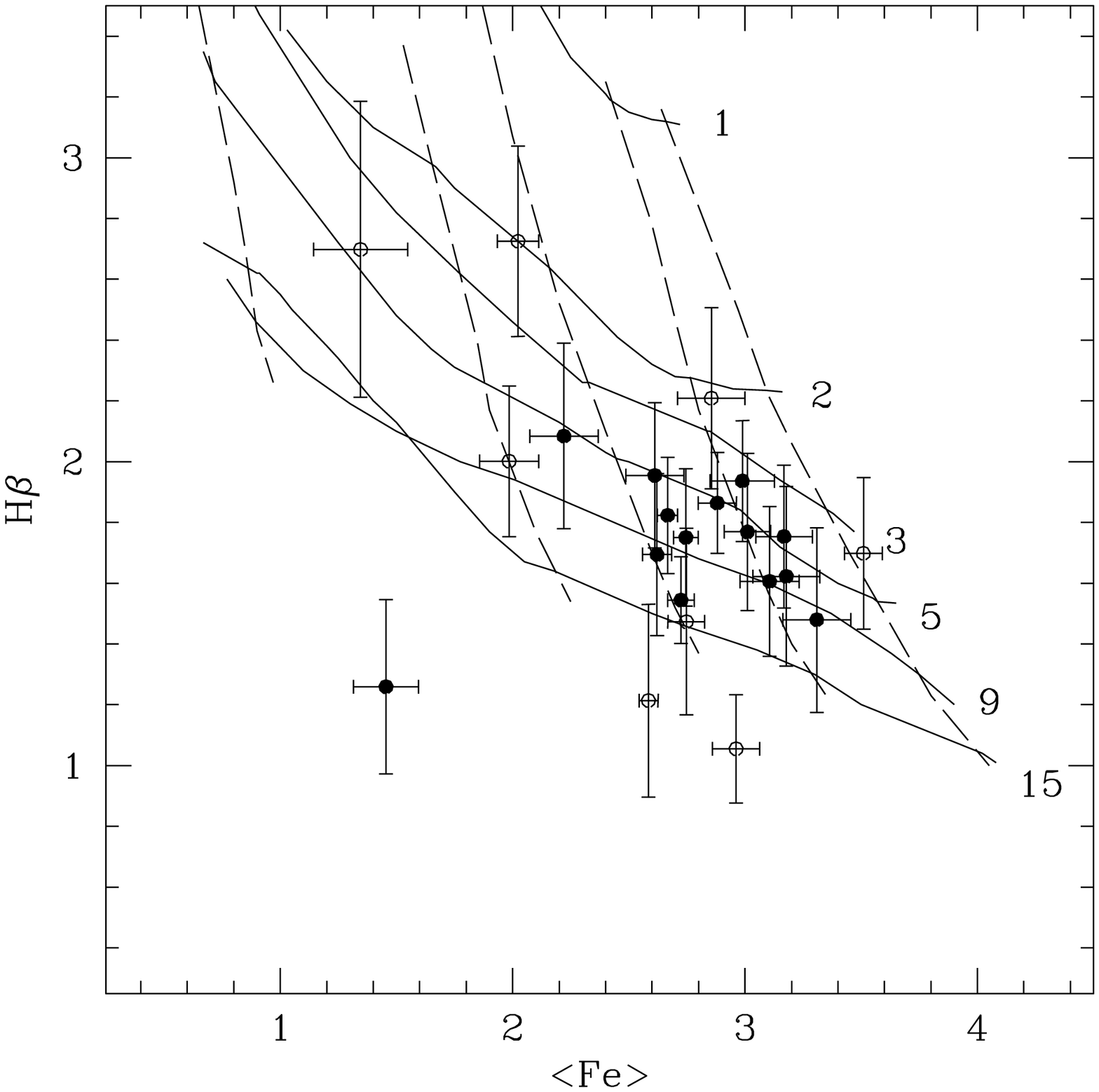}
\caption{\footnotesize Positions of 22 VGs
with their error bars. Shown are curves of 
constant age in Gyr (solid curves) and constant metal abundance, $[Z/H]$
(dashed curves: $[Z/H]$ = 0.67, 0.35, 0.0, -0.33, -1.35, from right to
left, interpolated from TMB models for
$[\alpha/H]=0.15$. Objects with detected H$\beta$ emission are open circles.
}
\label{fehbeta}
\end{figure}

\clearpage
\begin{figure}
\centering
\includegraphics[height=18truecm]{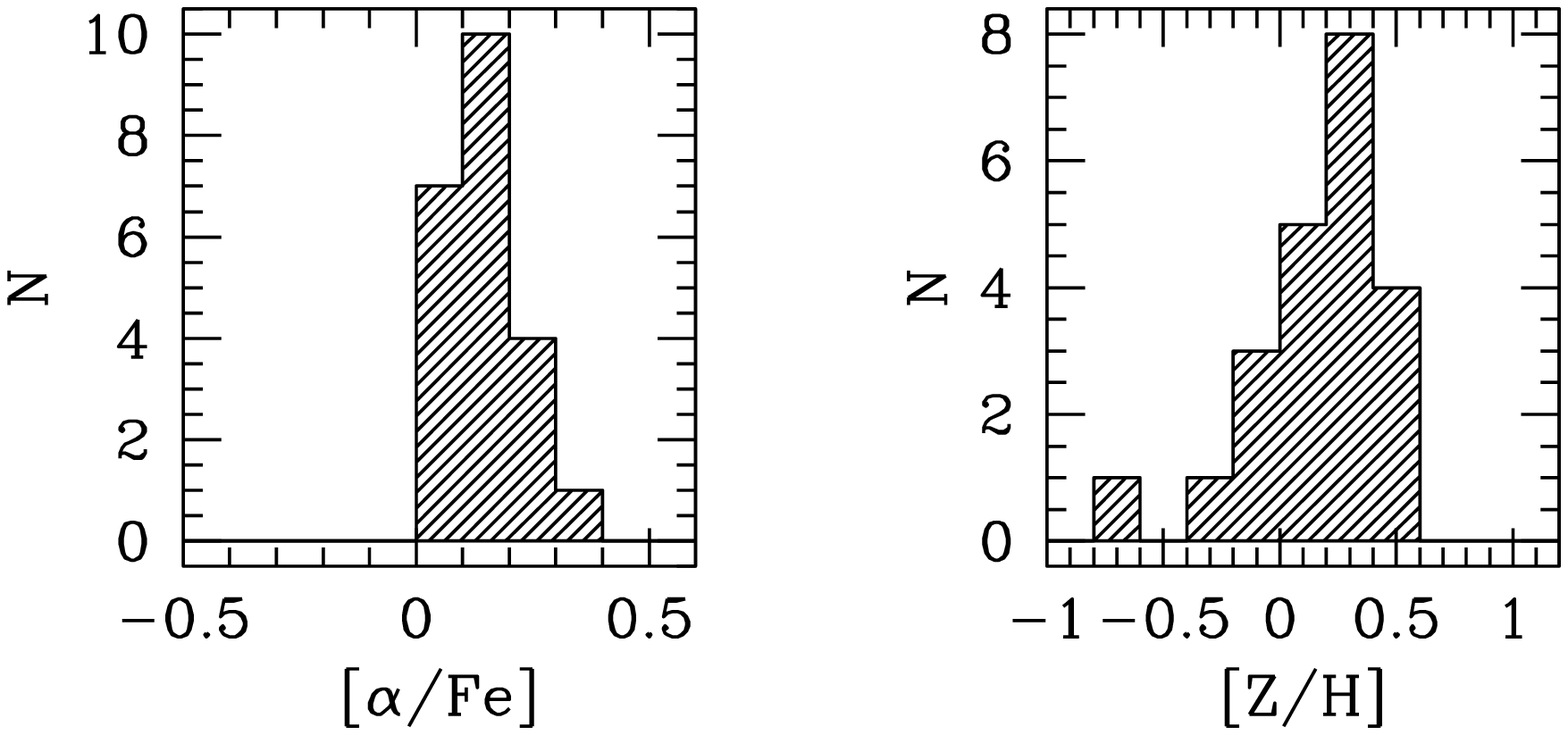}
\caption{Histograms of $[Z/H]$ and $[\alpha/Fe]$ for the program galaxies 
derived from (Mg$_b$ - $<{\rm{Fe}}>$) diagrams using the TMB models and 
the derived ages of the objects as described in the text.}
\end{figure}

\clearpage
\begin{figure}
\centering
\includegraphics[height=9truecm]{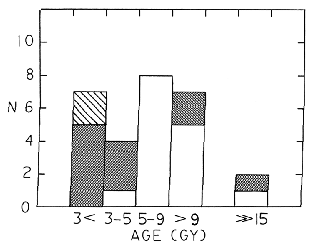}
\caption{\footnotesize
Histogram of the ages of galaxies read from Figure 8. Filled
bars represent galaxies observed with H$\beta$ emission.
Open bars denote objects for which no H$\beta$ emission
component was detected. The $\gg 15$ column denotes objects that
lie outside the grid in Figure 8. They are added at the top of
the $3<$ column as the hatched section following the discussion in the text.
}
\label{ageshisto}
\end{figure}

\clearpage
\begin{figure}
\centering
\includegraphics[height=18truecm]{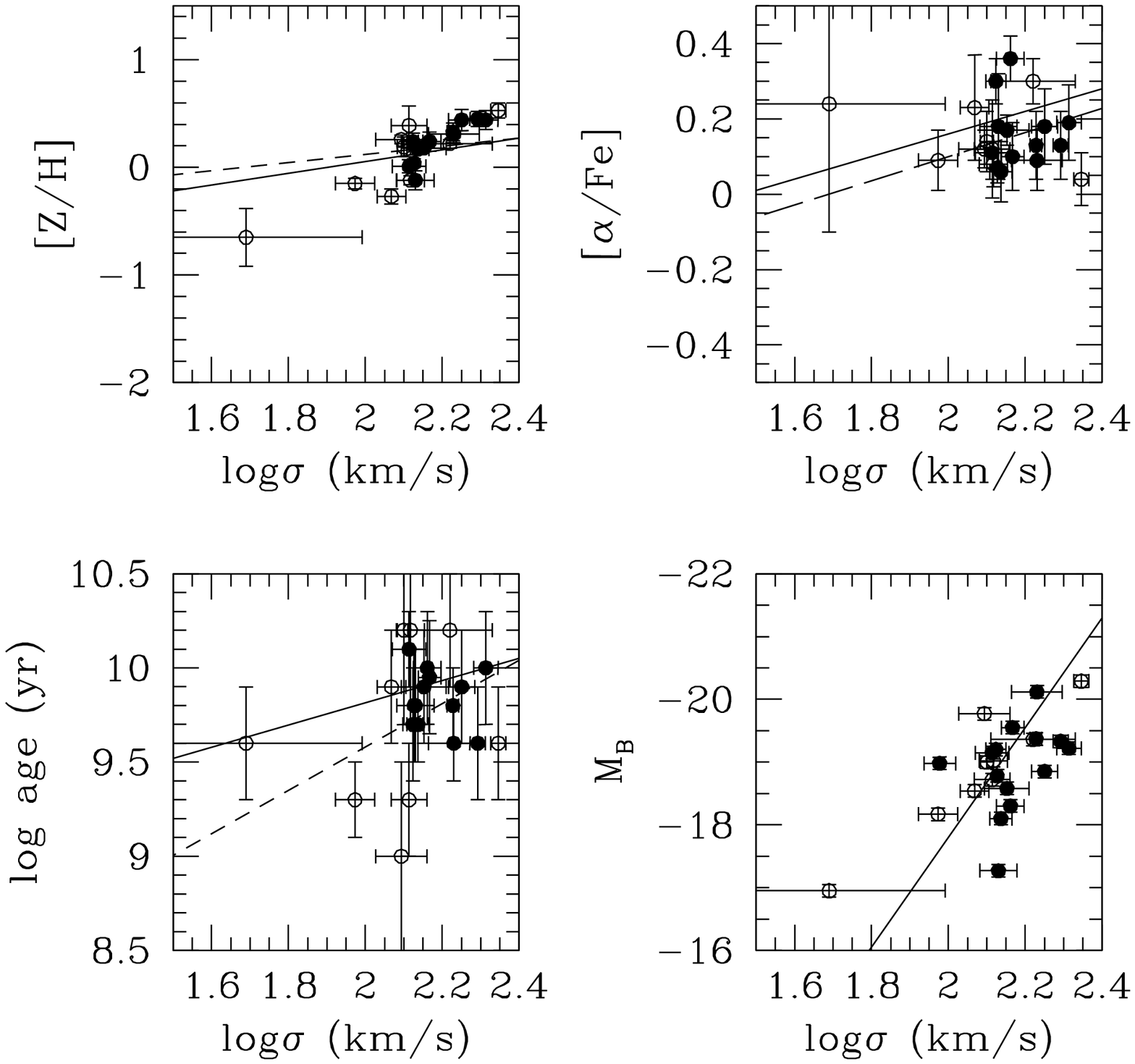}
\caption{\footnotesize
The metal abundances, ages, and absolute $B$ magnitudes of the VGs plotted
against available velocity dispersions. The linear relations for the
cluster sample of Nelan \etal (2005) are the solid lines and those
of the Bernardi \etal (2006) study are the dashed.
The line in the ($M_B, \log\sigma$ plot is the Faber-Jackson relation of the 
Coma Cluster (Dressler \etal 1987). Objects with detected H$\beta$ emission 
are open circles.  
}
\label{sigmaplots}
\end{figure}

\clearpage
\begin{figure}
\centering
\includegraphics[height=16truecm]{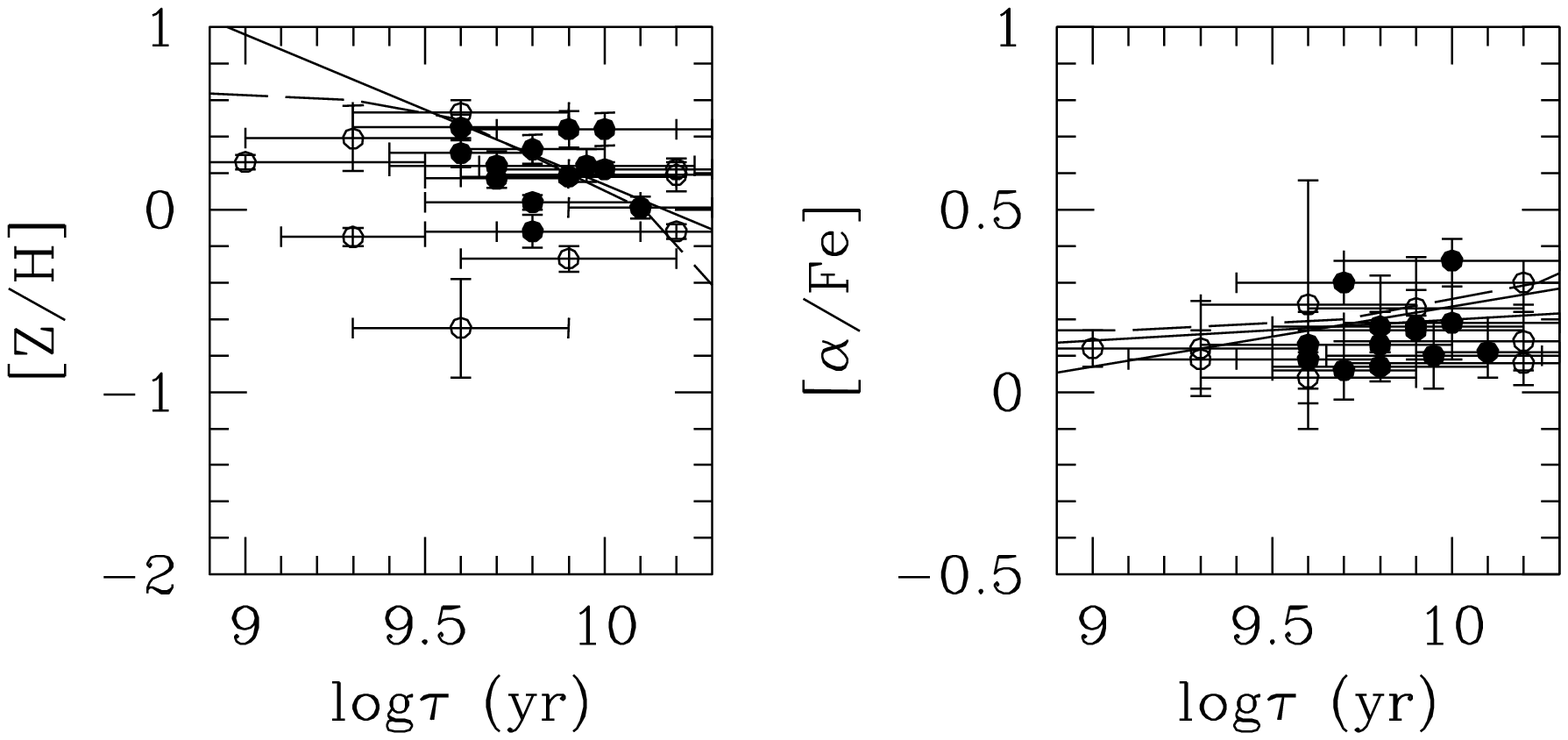}
\caption{\footnotesize
The metal abundances, [$\alpha$/Fe] and [Z/H] plotted against logarithm of
age, $\tau$ for galaxies with ages in Table 4. 
The solid lines are the relations given in Bernardi \etal (2006) and the
dashed curves are from Mehlert \etal (2003) for the Coma cluster.
Objects with detected H$\beta$ emission are open circles.
}
\label{metalages}
\end{figure}

\clearpage 
\begin{figure}
\centering
\includegraphics[height=10truecm]{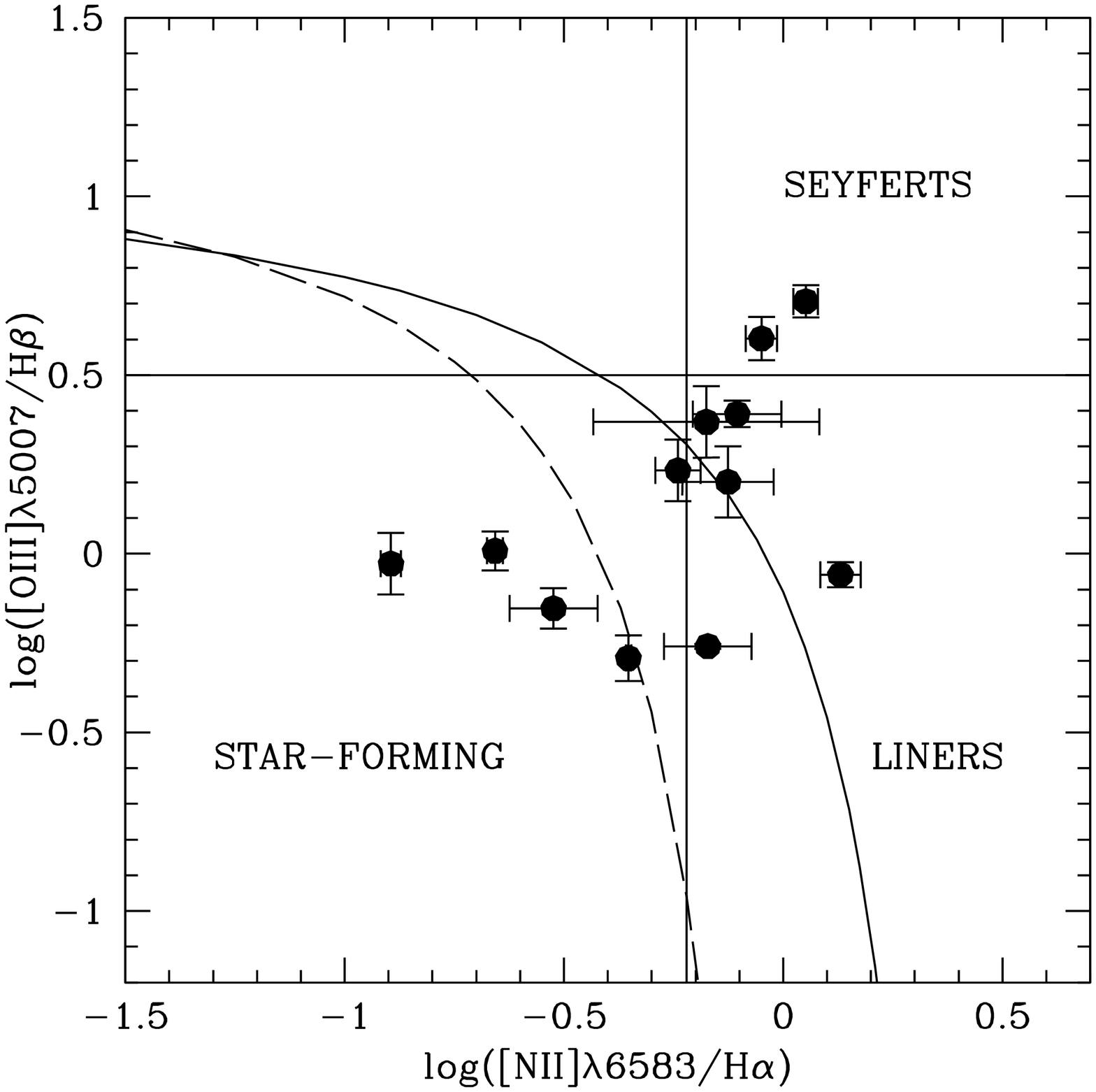}
\caption{\footnotesize Diagnostic diagram showing the positions of the 
galaxies in Table 3 with error bars. The line separating the Seyfert and liner
regions is that given in Yan et al. (2006). The 
solid and dashed curves demarcating the
edge of the star-forming region are taken from Kewley et al. (2001) and
Kaufmann et al. (2003).
}
\end{figure}

\end{document}

%% file: tab2.tex
  1   &1    & 7859& 14&  126&  5&   0.736&0.178&   0.319&   4.442 &0.174&   3.104&0.120&   2.819&0.078\\            
  2  &2    & 7059& 14&  130& 13&   1.241&0.143&   0.303&   3.958 &0.193&   3.022&0.080&   2.428 & 0.013\\              
  3        &2    & 6325& 10&  134&  2&   1.832&0.191&  - 0.001&   3.601 &0.022&   2.846&0.058&   2.487&0.018\\              
  4        &2    & 6255& 05&  170& 26&   1.366&0.199&   0.569&   3.863 &0.035&   3.228&0.193&   2.749&0.038\\              
  5    &2    & 5930& 11&  137&  9&   1.846&0.166&   0.018&   3.847 &0.198&   3.030&0.058&   2.731&0.101\\              
  6    &1    & 6363& 20&  147& 10&   1.700&0.246&  - 0.094&   4.455 &0.221&   3.329&0.151&   2.882&0.096\\              
  7    &1    & 6910& 26&  206& 15&   1.307&0.304&   0.172&   5.157 &0.258&   3.250&0.175&   3.367&0.110\\              
  8    &1    & 6267& 20&  117& 10&   1.264&0.248&   0.737&   2.831 &0.222&   2.398&0.152&   1.572&0.096\\              
  9    &1    & 6027& 26&  130& 14&   0.505&0.298&  1.724&   3.641 &0.255&   2.859&0.173&   2.851&0.109 \\              
 10    &1    & 6202& 25&  178& 14&   1.082&0.296&   0.540&   4.709 &0.253&   3.214&0.172&   3.141&0.108\\              
 11   &1    & 5565& 47&   49& 34&  -2.125&0.487&  4.823&   2.069 &0.367&   0.989&0.244&   1.703&0.149 \\              
 12   &2    & 5594& 22&  131& 11&  -2.038&0.317&  3.252&   3.427 &0.144&   2.772&0.057&   2.398&0.010\\              
 13   &1    & 6377& 19&  196&  9&   1.204&0.235&  0.549&   4.336 &0.214&   3.459&0.146&   2.877&0.093\\              
 14   &2    & 6090& 01&  145& 12&   1.486&0.266&   0.208&   5.046 &0.054&   2.687&0.010&   2.556&0.089 \\              
 15   &2    & 7318& 08&  124& 19&   -0.300&0.259&  4.162&  2.571 &0.060&   2.461&0.055&   2.068&0.024\\              
 16   &2    & 7695& 02&  169&  6&   1.585&0.258&   0.184&   4.335 &0.193&   3.180&0.050&   2.840&0.132 \\              
 17   &2    & 5158& 10&   94& 11&   -0.459&0.313&  3.184&  2.351 &0.014&   2.109&0.114&   1.937&0.054 \\              
 18      &2    & 5778& 02&  133&  8&   1.868&0.239&   0.086&   4.564 &0.019&   2.897&0.043&   2.326&0.172\\              
 19   &2    & 5407& 05&  142& 19&   1.592&0.226&   0.158&   4.152 &0.020&   2.862&0.005&   2.629&0.075 \\              
 20   &1    & 6548& 26&  135& 15&   1.804&0.305&   0.280&   3.362 &0.260&   1.737&0.176&   2.705&0.110 \\              
 21   &2    & 5067& 08&    -&  -&  -9.021&0.352&  3.219&   1.652 &0.012&   1.102&0.288&   1.178&0.028 \\              
 22   &1    & 5253& 93&  144& 28&   -7.123&0.857& 18.971&  1.062 &0.557&   0.423&0.364&   0.662&0.215 \\
 23     &2    & 4880& 02&  166& 42&   0.021&0.307&  1.452&   4.944 &0.077&   2.890&0.045&   2.604&0.102\\
 24      &2    & 5076& 06&  222& 10&   1.064&0.249&  0.634&   4.437 &0.292&   3.106&0.049&   3.913&0.105\\
 25    &1    & 7911&101&    -& - &  -1.924&0.921&  10.570&   0.311 &0.587&   1.182&0.382&  -0.131&0.225 \\  
 26    &1    & 6922& 24&   95&  9&   0.538&0.287&  0.721 &   0.435 &0.248&   1.971&0.168&   0.939&0.106    \\

%% file: tab3.tex
01& +0.886 &-0.958  &-0.175 &-0.287  &-0.671 &+0.369 \\
02& +0.956 &-0.946  &   -    &+1.448  &   -    &   -    \\
03& +2.361 &  -     &   -    &+2.069  &  -     &  -     \\
04& -0.245 &-2.245  &   -    &+3.123  &  -     &  -     \\
05& +2.241 &  -     &   -    &+2.356  &  -     &  -     \\
06& +2.745 &  -     &   -    &+3.162  &  -     &  -     \\
07& +1.545 &  -     &  -     &+2.145  &  -     &  -     \\
08& -0.996 &-2.483  &-0.125 &-0.381  &-0.605 &+0.201 \\
09& -5.346 &-6.021  &-0.105 &-0.987  &-2.246 &+0.391 \\
10& -0.112 &-1.494  &   -    &+2.545  &   -    &  -     \\
11& -19.385&-6.500  &-0.524 &-6.483  &-4.554 &-0.153 \\
12& -12.312&-9.850  &-0.172 &-4.168  &-2.295 &-0.259 \\
13& -0.151 &-0.979  &   -    &+2.073  &  -    &   -    \\
14& +1.383 &  -     &  -     &+3.073  &  -     &  -     \\
15& -16.410&-8.310  &-0.353 &-2.985  &-1.522 &-0.292 \\
16& +1.493 &  -     &   -    &+2.098  &   -    &  -     \\
17& -12.008&-8.245  &-0.240 &-2.194  &-3.753 &+0.233 \\
18& +1.931 &  -     &   -    &+2.881  &   -    &  -     \\
19& +1.608 &  -     &   -    &+2.112  &  -     &  -     \\
20& +1.058 &-0.614  &  -     &+1.636  &  -     &  -     \\
21& -12.164&-19.570 &+0.131 &-14.502 &-12.656&-0.059 \\
22& -83.050&-10.885 &-0.894 &-19.615 &-18.398&-0.028 \\
23& -4.215 &-7.365  &+0.052 &-0.717  &-3.645 &+0.706 \\
24& -0.533 &-2.545  &-0.050 &-0.240  &-0.962 &+0.602 \\
25& -45.25 &-11.265 &-0.625 &-3.976  &-4.050 &+0.008 \\
26& -0.924 &-1.328  &   -    &+0.995  &   -    &   -    \\